# Parameter Convergence Radar Detector Based on VAMP Deep Unfolding

Haoyun Zhang, Jianghong Han, Xueqian Wang, Gang Li*, *Senior Member, IEEE*, Xiao-Ping Zhang, *Fellow, IEEE*

***Abstract*—Compared with the sparse recovery process in traditional compressed sensing (CS) radar detector CAMP, vector AMP deep unfolding (VAMP-DU) can achieve sparse recovery over a broader range of observation matrices, with faster convergence speed and higher recovery accuracy. However, the distribution of the error term in VAMP-DU remains unknown, which renders the distribution of the test statistic in CS radar detection undetermined and thus hinders threshold setting under a given false alarm rate when VAMP-DU is applied to CS radar detection. In this work, we theoretically prove that the error term in VAMP-DU follows a Gaussian distribution by leveraging a general state evolution (SE). Based on the Gaussianity, we propose a new parameter convergence radar detector (PCRD) as the CS detector to calculate the distribution parameter of the test statistic and realize target detection under a given false alarm rate. Specifically, PCRD exploits the Gaussian property of error term in VAMP-DU to exhibit superior false alarm control capability, while leveraging the improved recovery accuracy of VAMP-DU to further enhance target detection performance. Numerical simulations validate the Gaussianity of the error term in VAMP-DU and show the superiority of the VAMP-DU-based PCRD over existing approaches in both false alarm control accuracy and target detection performance.**

## I. INTRODUCTION

In recent years, compressed sensing (CS) radar has achieved rapid development [2–9]. On the one hand, by exploiting the sparsity of signals, CS radar systems perform radar signal processing with fewer measurements than those required by Nyquist sampling, thus reducing resource consumption in multiple dimensions such as the time and frequency domains [10–17]. On the other hand, CS radar can address the inherent sidelobe issues of traditional radar systems to enhance radar target detection performance [18–21].

In CS radar target detection, threshold setting under a given false alarm rate is of critical importance. Several detectors have been proposed to achieve target detection under a given false alarm rate based on CS radar. To enable threshold setting under a given false alarm rate, the distribution of the error term $\mathbf{w} \in \mathbb{C}^{N\times 1}$ must be known in advance [22]. This prior knowledge of the error term's distribution allows the distribution of the test statistic $\mathbf{r} \in \mathbb{C}^{N\times 1}$ to be determined. According to the approach of test statistic acquisition, these CS radar detectors are categorized into two types, which will be elaborated on subsequently.

The first type of CS radar detectors utilizes the final sparse solution of the sparse recovery algorithm to generate a test statistic with a specific distribution. A typical detector in this category is the complex row-orthogonal debiased detector (CROD) [22] proposed by Na et al. CROD can set threshold when the observation matrix is row-orthogonal, which can be applied to CS radar in some practical scenarios [23–27]. However, CROD relies on prior knowledge of the additive white Gaussian noise (AWGN) power. In non-stationary environments, achieving real-time and accurate estimation of the AWGN power poses challenges. SDL [28] can be utilized as a detector of the first type, which does not require AWGN power previously. However, SDL can only be applied to CS radar detection problems with Gaussian random matrices. Moreover, the detectors of the first type fail to fully exploit the distribution property of the pseudo-measurement in certain sparse recovery algorithms, which is an intermediate result in such algorithms, thus the threshold-setting accuracy can be improved.

The second type of CS radar detectors utilizes the pseudo-measurement with a specific distribution from approximate message passing (AMP)-based sparse recovery algorithms as the test statistic. By leveraging the Gaussian distributed pseudo-measurement as the test statistic, Anitori et al. proposed complex AMP (CAMP) [29–33] to obtain a closed-form expression describing the relationship between the threshold and the false alarm rate. However, theoretically, CAMP is only applicable to CS radar detection problems with Gaussian random observation matrices. In practical CS radar systems, the prior requirements on the observation matrices imposed by CAMP cannot be satisfied, which decreases the sparse recovery accuracy and detection performance. Therefore, it is necessary to employ sparse recovery algorithm with pseudo-measurements featuring a specific distribution which can be applied to a broad range of observation matrix scenarios.

Vector AMP deep unfolding (VAMP-DU) [34] is an AMP-based sparse recovery algorithm offering a novel perspective for addressing this challenge. Utilizing the deep unfolding approach, the parameters in traditional VAMP [35] can be modified to achieve greater sparse recovery accuracy and faster convergence under a broad range of observation matrix conditions. However, parameter modifications in deep unfolding render the state evolution (SE) in traditional VAMP invalid. Therefore, the distribution of the error term $\mathbf{w}$ in VAMP-DU is unknown. Consequently, the distribution of the pseudo-measurement $\mathbf{r}$ in

This work is supported by the National Natural Science Foundation of China under Grant 62388102 and 62341130, with a preliminary version presented as a conference paper [1].

VAMP-DU, which should serve as the test statistic, cannot be determined, thus hindering threshold setting under a given false alarm rate in the application of VAMP-DU to CS radar detection.

In this paper, we theoretically investigate the distribution of the error term $\mathbf{w}$ in VAMP-DU. Based on the distribution of the error term, we propose a parameter convergence radar detector (PCRD) as a CS radar detector to estimate the distribution parameter of the test statistic $\mathbf{r}$ and implement target detection. Numerical simulations demonstrate that VAMP-DU-based PCRD achieves more accurate false alarm control and superior target detection performance compared to state-of-the-art approaches.

The contributions of this paper are as follows. First, the distribution of the error term $\mathbf{w}$ obtained by the VAMP-DU algorithm is investigated. Through our theoretical proof in the view of more general SE, it is concluded that the components of $\mathbf{w}$ in VAMP-DU follow a complex Gaussian distribution, denoted as $w_i \sim \mathcal{CN}(0, 2\sigma_{\mathrm{R,I}}^2)$. However, the variance $\sigma_{\mathrm{R,I}}^2$ cannot be obtained directly. To our knowledge, the exploration of the distribution of the error term $\mathbf{w}$ in VAMP-DU is novel.

Second, we propose PCRD based on VAMP-DU to estimate the distribution parameter $\sigma_{\mathrm{R,I}}^2$ and implement target detection. PCRD exploits the Gaussianity of the error term to achieve more accurate false alarm control, while utilizing the improved accuracy in VAMP-DU to enhance target detection performance, which bridges the gap between the advanced data-driven learning approach and CS radar target detection under a given false alarm rate in practical scenarios.

The organization of this paper is as follows: In Section II, the problem formulation in CS radar detection based on VAMP-DU is briefly presented. In Section III, we theoretically prove the Gaussianity of pseudo-measurement from the perspective of a more general SE. In Section IV, we propose PCRD, along with an analysis of its effectiveness of PCRD in terms of variance estimation. In Section V, through numerical simulations, the Gaussianity of error term is validated, and the superiority of VAMP-DU-based PCRD in terms of false alarm control and target detection is evaluated. The conclusion of the paper is given in Section VI.

## II. Problem Formulation

The CS radar target detection aims to detect targets in the sparse observation scene $\mathbf{x}_0 \in \mathbb{C}^{N\times 1}$ element-wisely from the received sampled signal $\mathbf{y} \in \mathbb{C}^{M\times 1}$. The signal model is given by:

$$\mathbf{y} = \mathbf{A}\mathbf{x}_0 + \mathbf{n}, \tag{1}$$

where $\mathbf{A}\in\mathbb{C}^{M\times N}\,(M<N)$ represents the complex observation matrix, containing the information of the transmitting waveform and observation geometry; $\mathbf{n}\in\mathbb{C}^{N\times 1}$ represents the complex AWGN with independent and identically distributed (i.i.d.) components $n_i$ following a complex Gaussian distribution, i.e. $n_i \sim \mathcal{CN}(0, 2\sigma^2)$. The sparsity of $\mathbf{x}_0$ is denoted by $L_0 = \|\mathbf{x}_0\|_0 = |\mathrm{supp}(\mathbf{x}_0)|$, where $\|\mathbf{x}_0\|_0$ and $\mathrm{supp}(\mathbf{x}_0)$ represent the $l_0$ norm and the support set of $\mathbf{x}_0$, respectively, and $|\mathrm{supp}(\mathbf{x}_0)|$ represents the cardinality of $\mathrm{supp}(\mathbf{x}_0)$. The signal density $\rho$ is defined as $\rho = L_0 / N$.

The complex-valued CS radar signal model in Equation (1) should be decomposed into a real-valued formulation to adapt to the deep unfolding network [36], expressed as:

$$\mathbf{y}_{\mathrm{R,I}} = \mathbf{A}_{\mathrm{R,I}}\mathbf{x}_{0,\mathrm{R,I}} + \mathbf{n}_{\mathrm{R,I}}, \tag{2}$$

where

$$\mathbf{y}_{\mathrm{R,I}} = \begin{bmatrix}\mathrm{Re}(\mathbf{y})\\ \mathrm{Im}(\mathbf{y})\end{bmatrix} = \begin{bmatrix}\mathbf{y}_{\mathrm{R}}\\ \mathbf{y}_{\mathrm{I}}\end{bmatrix}, \mathbf{A}_{\mathrm{R,I}} = \begin{bmatrix}\mathrm{Re}(\mathbf{A}) & -\mathrm{Im}(\mathbf{A})\\ \mathrm{Im}(\mathbf{A}) & \mathrm{Re}(\mathbf{A})\end{bmatrix} = \begin{bmatrix}\mathbf{A}_{\mathrm{R}} & -\mathbf{A}_{\mathrm{I}}\\ \mathbf{A}_{\mathrm{I}} & \mathbf{A}_{\mathrm{R}}\end{bmatrix},$$
$$\mathbf{x}_{0,\mathrm{R,I}} = \begin{bmatrix}\mathrm{Re}(\mathbf{x}_0)\\ \mathrm{Im}(\mathbf{x}_0)\end{bmatrix} = \begin{bmatrix}\mathbf{x}_{0,\mathrm{R}}\\ \mathbf{x}_{0,\mathrm{I}}\end{bmatrix}, \mathbf{n}_{\mathrm{R,I}} = \begin{bmatrix}\mathrm{Re}(\mathbf{n})\\ \mathrm{Im}(\mathbf{n})\end{bmatrix} = \begin{bmatrix}\mathbf{n}_{\mathrm{R}}\\ \mathbf{n}_{\mathrm{I}}\end{bmatrix} \tag{3}$$

with $\mathrm{Re}(\cdot)$ and $\mathrm{Im}(\cdot)$ representing the operations of extracting the real and the imaginary parts from complex values, respectively. Here, $\mathbf{y}_{\mathrm{R}} = \mathrm{Re}(\mathbf{y})$ and $\mathbf{y}_{\mathrm{I}} = \mathrm{Im}(\mathbf{y})$ represent the real and imaginary parts of the complex vector $\mathbf{y}$, respectively, and $\mathbf{y}_{\mathrm{R,I}}$ represents the real-valued formulation of $\mathbf{y}$. Similarly, the real-valued formulation of $\mathbf{A}$, $\mathbf{x}_0$ and $\mathbf{n}$ can be defined as $\mathbf{A}_{\mathrm{R,I}}$, $\mathbf{x}_{0,\mathrm{R,I}}$ and $\mathbf{n}_{\mathrm{R,I}}$, respectively, in Equation (3). In CS radar, the sparse recovery process using VAMP-DU [34] aims to obtain the sparse solution and pseudo-measurement. (VAMP-DU is briefly reviewed in Appendix A.) The sparse recovery process using VAMP-DU can be expressed as:

$$\hat{\mathbf{x}}_{\mathrm{R,I}}, \mathbf{r}_{\mathrm{R,I}} = \mathbf{F}_{\mathrm{VAMP\text{-}DU}}(\mathbf{y}_{\mathrm{R,I}}, \mathbf{A}_{\mathrm{R,I}}), \tag{4}$$

where $\hat{\mathbf{x}}_{\mathrm{R,I}}$ and $\mathbf{r}_{\mathrm{R,I}}$ represent the sparse solution and pseudo-measurement in real-valued formulation, respectively. The sparse solution and pseudo-measurement can be represented as:

$$\begin{aligned}\hat{\mathbf{x}}_0 &= \hat{\mathbf{x}}_{\mathrm{R}} + j\hat{\mathbf{x}}_{\mathrm{I}} = \mathbf{f}_{\mathrm{R}}(\hat{\mathbf{x}}_{\mathrm{R,I}}) + j\mathbf{f}_{\mathrm{I}}(\hat{\mathbf{x}}_{\mathrm{R,I}}),\\ \mathbf{r} &= \mathbf{r}_{\mathrm{R}} + j\mathbf{r}_{\mathrm{I}} = \mathbf{f}_{\mathrm{R}}(\mathbf{r}_{\mathrm{R,I}}) + j\mathbf{f}_{\mathrm{I}}(\mathbf{r}_{\mathrm{R,I}}),\end{aligned} \tag{5}$$

respectively, where $j$ is the imaginary unit and the functions $\mathbf{f}_{\mathrm{R}}(\cdot)$ and $\mathbf{f}_{\mathrm{I}}(\cdot)$ are defined as:

$$\begin{aligned}\mathbf{f}_{\mathrm{R}}(\mathbf{h}_{\mathrm{R,I}}) &= \mathbf{h}_{\mathrm{R}},\\ \mathbf{f}_{\mathrm{I}}(\mathbf{h}_{\mathrm{R,I}}) &= \mathbf{h}_{\mathrm{I}},\end{aligned} \tag{6}$$

respectively. Here, the function $\mathbf{f}_{\mathrm{R}}(\mathbf{h}_{\mathrm{R,I}})$ obtains the real part $\mathbf{h}_{\mathrm{R}}$ from the real-valued formulation vector $\mathbf{h}_{\mathrm{R,I}}$. Similarly, the function $\mathbf{f}_{\mathrm{I}}(\mathbf{h}_{\mathrm{R,I}})$ obtains the imaginary part $\mathbf{h}_{\mathrm{I}}$ from the real-valued formulation $\mathbf{h}_{\mathrm{R,I}}$.

For radar target detection, there are two hypotheses defined as follows:

$$\begin{cases} H_{0,i} : x_{0,i} = 0 \\ H_{1,i} : x_{0,i} \neq 0 \end{cases}, \text{ for } i = 1, 2, \dots, N, \tag{7}$$

where $x_{0,i}$ represents the $i$-th element of the observation scene $\mathbf{x}_0$. Here, hypothesis $H_{0,i}$ indicates that no target exists in the $i$-th element of the sparse scene $\mathbf{x}_0$, i.e. $x_{0,i}=0$; hypothesis $H_{1,i}$ indicates that a target exists in the $i$-th element of $\mathbf{x}_0$, i.e. $x_{0,i}\neq 0$.

In this paper, the pseudo-measurement $\mathbf{r}$ obtained by VAMP-DU is selected as the test statistic, since the pseudo-measurement $\mathbf{r}$ can be regarded as the observation scene $\mathbf{x}_0$ disturbed by error term $\mathbf{w}$ with a specific distribution, given by:

$$\mathbf{r}=\mathbf{x}_0+\mathbf{w}, \tag{8}$$

which is requisite for target detection under a given false alarm rate. According to the Neyman-Pearson criterion, the decision criterion can be derived as:

$$\frac{p\left(|r_i| \mid H_{1,i}\right)}{p\left(|r_i| \mid H_{0,i}\right)} \underset{H_1}{\overset{H_0}{\lessgtr}} -\lambda_{P_{\mathrm{fa}}}, \tag{9}$$

where $|r_i|$ is the $i$-th element of $|\mathbf{r}|$, $p(|r_i| \mid H_{1,i})$ and $p(|r_i| \mid H_{0,i})$ are the probability density functions (PDFs) of $|r_i|$ under the hypotheses $H_{1,i}$ and $H_{0,i}$, respectively; $-\lambda_{P_{\mathrm{fa}}}$ denotes the detection threshold of the likelihood ratio corresponding to a certain false alarm rate $P_{\mathrm{fa}}$. It is obvious that the design of $-\lambda_{P_{\mathrm{fa}}}$ necessitates priori knowledge of the test statistic distribution $p(|r_i| \mid H_{1,i})$ and $p(|r_i| \mid H_{0,i})$. However, the distribution of the pseudo-measurement element $|r_i|$ in VAMP-DU has not been investigated. Consequently, radar detection under a given false alarm rate cannot be performed directly. To address this issue, we will explore the distribution of the error term $\mathbf{w}=\mathbf{r}-\mathbf{x}_0$ to obtain $p(|r_i| \mid H_{1,i})$ and $p(|r_i| \mid H_{0,i})$, and propose PCRD as a CS radar detector based on VAMP-DU.

## III. Error Term Distribution Proof

We prove the distribution of the error term $\mathbf{w}$, formalized by the following theorem:

*Theorem 1*. The components $w_i$ of the error term $\mathbf{w}$ follow a complex Gaussian distribution $w_i \sim \mathcal{CN}(0, 2\sigma_{\mathrm{R,I}}^2)$.

*Proof.* To prove *Theorem 1*, we proceed in two steps:

Step 1: We propose a general recursion and prove that for any iteration $k$, the components $q_{kn}$ of the variable $\mathbf{q}_k$ in the recursion converge empirically with second order moments to $Q_k$, where $Q_k$ is a Gaussian random variable (The definition of empirical convergence with $p-th$ order moments can be found in [35]).

Step 2: We derive that the VAMP-DU is a special case of the general recursion, and the error term in real-valued formulation $\left[\mathrm{Re}(\mathbf{w}) \;\; \mathrm{Im}(\mathbf{w})\right]^{\top}=\mathbf{w}_{\mathrm{R,I}}=\mathbf{r}_{\mathrm{R,I}}-\mathbf{x}_{0,\mathrm{R,I}}$ is a special case of $\mathbf{q}_{k=K_{\mathrm{it}}}$, where $K_{\mathrm{it}}$ is the number of iterations in VAMP-DU.

Thus, the components $w_{\mathrm{R,I},i}$ of $\mathbf{w}_{\mathrm{R,I}}$ follow a Gaussian distribution, represented as $w_{\mathrm{R,I},i} \sim \mathcal{N}(0,\sigma_{\mathrm{R,I}}^2)$, and accordingly, the components $w_i$ of the error term $\mathbf{w}=\mathbf{w}_{\mathrm{R}}+j\mathbf{w}_{\mathrm{I}}=\mathbf{f}_{\mathrm{R}}\left(\mathbf{w}_{\mathrm{R,I}}\right)+j\mathbf{f}_{\mathrm{I}}\left(\mathbf{w}_{\mathrm{R,I}}\right)$ follow a complex Gaussian distribution, that is $w_i \sim \mathcal{CN}(0,2\sigma_{\mathrm{R,I}}^2)$.

We present Step 1 and Step 2 of the proof in the following two subsections.

### A. Step 1: A General Recursion

We consider a more general recursion than that in [35]. The notation $\mathbf{V}' \in \mathbb{R}^{2N\times 2N}$ denotes an orthogonal matrix and $\mathbf{u}_0 \in \mathbb{R}^{2N}$ denotes a vector, both of which are given initially. The disturbance vectors can be defined as:

$$\mathbf{w}^p=\left(w_1^p,\ldots,w_n^p\right), \quad \mathbf{w}^q=\left(w_1^q,\ldots,w_n^q\right), \tag{10}$$

where the components $w_n^p \in \mathbb{R}^{n_p}$ and $w_n^q \in \mathbb{R}^{n_q}$. Note that $n_p$ and $n_q$ do not vary with $N$. Then the recursion is given as:

$$\mathbf{p}_k=\mathbf{V}'\mathbf{u}_k \tag{11a}$$

$$\alpha_{1k}=\left\langle \mathbf{f}_{p,k}'\left(\mathbf{p}_k,\mathbf{w}^p,\gamma_{1k}\right)\right\rangle, \quad \gamma_{2k}=\Gamma_1\left(\gamma_{1k},\alpha_{1k}\right) \tag{11b}$$

$$\mathbf{v}_k=C_1\left(\alpha_{1k}\right)\left[\mathbf{f}_{p,k}\left(\mathbf{p}_k,\mathbf{w}^p,\gamma_{1k}\right)-\alpha_{1k}\mathbf{p}_k\right] \tag{11c}$$

$$\mathbf{q}_k=\mathbf{V}'^{\top}\mathbf{v}_k \tag{11d}$$

$$\alpha_{2k}=\left\langle \mathbf{f}_{q,k}'\left(\mathbf{q}_k,\mathbf{w}^q,\gamma_{2k}\right)\right\rangle, \quad \gamma_{1,k+1}=\Gamma_2\left(\gamma_{2k},\alpha_{2k}\right) \tag{11e}$$

$$\mathbf{u}_{k+1}=C_2\left(\alpha_{2k}\right)\left[\mathbf{f}_{q,k}\left(\mathbf{q}_k,\mathbf{w}^q,\gamma_{2k}\right)-\alpha_{2k}\mathbf{q}_k\right]. \tag{11f}$$

In Recursion (11), the subscript $k$ denotes the index of iteration. Here, $\mathbf{f}_{p,k}(\cdot)$ and $\mathbf{f}_{q,k}(\cdot)$ are scalar-valued functions that vary with iterations, represented as:

$$\begin{aligned}\left[\mathbf{f}_{p,k}\left(\mathbf{p},\mathbf{w}^p,\gamma_1\right)\right]_n&=f_{p,k}\left(p_n,w_n^p,\gamma_1\right)\ \forall n,\\ \left[\mathbf{f}_{q,k}\left(\mathbf{q},\mathbf{w}^q,\gamma_2\right)\right]_n&=f_{q,k}\left(q_n,w_n^q,\gamma_2\right)\ \forall n,\end{aligned} \tag{12}$$

respectively. Variables $\alpha_{1k}$ and $\alpha_{2k}$ can be calculated as:

$$\alpha_{1k}=\left\langle \mathbf{f}_{p,k}'\left(\mathbf{p}_k,\mathbf{w}^p,\gamma_{1k}\right)\right\rangle=\frac{1}{N}\mathrm{Tr}\left[\frac{\partial \mathbf{f}_{p,k}\left(\mathbf{p}_k,\mathbf{w}^p,\gamma_{1k}\right)}{\partial \mathbf{p}_k}\right], \tag{13a}$$

$$\alpha_{2k}=\left\langle \mathbf{f}_{q,k}'\left(\mathbf{q}_k,\mathbf{w}^q,\gamma_{2k}\right)\right\rangle=\frac{1}{N}\mathrm{Tr}\left[\frac{\partial \mathbf{f}_{q,k}\left(\mathbf{q}_k,\mathbf{w}^q,\gamma_{2k}\right)}{\partial \mathbf{q}_k}\right], \tag{13b}$$

respectively. Note that Equation (12) is a generalization of the corresponding functions that do not vary with iterations in the recursion of [35]. Here, $f_{p,k}(\cdot)$ and $f_{q,k}(\cdot)$ are scalar-valued functions. The functions $\Gamma_i(\cdot)$ and $C_i(\cdot)$ are also scalar-valued. There exist $U_0$, $W^p$ and $W^q$ such that the components in the recursion converge empirically as:

$$\lim_{N\to\infty}\{u_{0n}\}_{n=1}^{N} \overset{PL(2)}{=} U_0, \tag{14}$$

and

$$\lim_{N\to\infty}\{w_n^p\}_{n=1}^{N} \overset{PL(2)}{=} W^p, \quad \lim_{N\to\infty}\{w_n^q\}_{n=1}^{N} \overset{PL(2)}{=} W^q. \tag{15}$$

Also, we have

$$\lim_{N\to\infty}\gamma_{10} = \bar{\gamma}_{10}, \tag{16}$$

for some $\bar{\gamma}_{10}$. The matrix $\mathbf{V}' \in \mathbb{R}^{2N\times 2N}$ is assumed to be uniformly distributed on the set of orthogonal matrices independent of $\bar{\gamma}_{10}$, $\mathbf{w}^p$ and $\mathbf{w}^q$.

Under the above assumptions, we define SE equations more general than those in [35] as:

$$\bar{\alpha}_{1k} = \mathbb{E}\left[f'_{p,k}\left(P_k, W^p, \bar{\gamma}_{1k}\right)\right] \tag{17a}$$

$$\tau_{2k} = C_1^2\left(\bar{\alpha}_{1k}\right)\left\{\mathbb{E}\left[f_{p,k}^2\left(P_k, W^p, \bar{\gamma}_{1k}\right)\right] - \bar{\alpha}_{1k}^2\tau_{1k}\right\} \tag{17b}$$

$$\bar{\gamma}_{2k} = \Gamma_1\left(\bar{\gamma}_{1k}, \bar{\alpha}_{1k}\right) \tag{17c}$$

$$\bar{\alpha}_{2k} = \mathbb{E}\left[f'_{q,k}\left(Q_k, W^q, \bar{\gamma}_{2k}\right)\right] \tag{17d}$$

$$\tau_{1,k+1} = C_2^2\left(\bar{\alpha}_{2k}\right)\left\{\mathbb{E}\left[f_{q,k}^2\left(Q_k, W^q, \bar{\gamma}_{2k}\right)\right] - \bar{\alpha}_{2k}^2\tau_{2k}\right\} \tag{17e}$$

$$\bar{\gamma}_{1,k+1} = \Gamma_2\left(\bar{\gamma}_{2k}, \bar{\alpha}_{2k}\right), \tag{17f}$$

which are initialized with

$$\tau_{10} = \mathbb{E}\left[U_0^2\right]. \tag{18}$$

Additionally, the random variables can be represented as:

$$P_k \sim \mathcal{N}\left(0, \tau_{1k}\right), \quad Q_k \sim \mathcal{N}\left(0, \tau_{2k}\right), \tag{19}$$

where $P_k$ and $Q_k$ are independent of $W^p$ and $W^q$.

According to *Lemma 1*, which we prove in Appendix B, the behavior of variables in Recursion (11) can be characterized by SE (17). Specifically, for any fixed iteration index $k$, almost surely the components of $\left(\mathbf{w}^q, \mathbf{q}_0, \ldots, \mathbf{q}_k\right)$ in Recursion (11) empirically converge as:

$$\lim_{N\to\infty}\left\{\left(w_n^q, q_{0n}, \ldots, q_{kn}\right)\right\}_{n=1}^{N} \overset{PL(2)}{=} \left(W^q, Q_0, \ldots, Q_k\right), \tag{20}$$

where $W^q$ is the random variable in Equation (15) and $\left(Q_0, \ldots, Q_k\right)$ in SE (17) is a Gaussian vector of zero mean, independent of $W^q$. Therefore, we can derive:

$$\lim_{N\to\infty}\{q_{kn}\}_{n=1}^{N} \overset{PL(2)}{=} Q_k, \tag{21}$$

where $Q_k$ is a Gaussian random variable of zero mean.

Thus, if we can prove that the error term in real-valued formulation $\mathbf{w}_{\mathrm{R,I}}$ is a special case of $\mathbf{q}_{k=K_{\mathrm{it}}}$, which will be elaborated on subsequently, we can then derive the components $w_{\mathrm{R,I},i}$ of $\mathbf{w}_{\mathrm{R,I}}$ follow a Gaussian distribution, and accordingly the components $w_i$ of the error term $\mathbf{w}$ follow a complex Gaussian distribution $w_i \sim \mathcal{CN}(0, 2\sigma_{\mathrm{R,I}}^2)$.

### *B. Step 2: A Special Case of the Recursion*

Next, we will rewrite the algorithm of VAMP-DU (Appendix A) to show that VAMP-DU is a special case of the Recursion (11) and $\mathbf{w}_{\mathrm{R,I}}$ is a special case of $\mathbf{q}_{k=K_{\mathrm{it}}}$. We implement singular-value-decomposition of observation matrix $\mathbf{A}_{\mathrm{R,I}} \in \mathbb{R}^{2M\times 2N}$, which can be represented as:

$$\mathbf{A}_{\mathrm{R,I}} = \mathbf{U}\mathrm{Diag}(\mathbf{s})\mathbf{V}^\top. \tag{22}$$

We define the error terms as:

$$\mathbf{u}_k = \mathbf{r}_{1k} - \mathbf{x}_{0,\mathrm{R,I}}, \quad \mathbf{q}_k = \mathbf{r}_{2k} - \mathbf{x}_{0,\mathrm{R,I}}. \tag{23}$$

The transform of which can be written as:

$$\mathbf{p}_k = \mathbf{V}^\top\mathbf{u}_k, \quad \mathbf{v}_k = \mathbf{V}^\top\mathbf{q}_k. \tag{24}$$

Next, we define

$$\xi = \mathbf{U}^\top\mathbf{n}_{\mathrm{R,I}}, \quad \mathbf{w}^p = (\xi, \mathbf{s}), \quad \mathbf{w}^q = \mathbf{x}_{0,\mathrm{R,I}}. \tag{25}$$

The functions which can be implemented element-wisely, are defined as:

$$f_{p,k}\left(p, (\xi, s), \gamma_1\right) = \frac{\gamma_{wk}s\xi + \gamma_1 p}{\gamma_{wk}s^2 + \gamma_1}, \tag{26a}$$

$$f_{q,k}\left(q, x_{0,\mathrm{R,I}}, \gamma_2\right) = g_{2,k}\left(q + x_{0,\mathrm{R,I}}, \gamma_2\right) - x_{0,\mathrm{R,I}}. \tag{26b}$$

The Recursion (11) can be written as:

$$\mathbf{p}_k = \mathbf{V}'\mathbf{u}_k \tag{27a}$$

$$\alpha_{1k} = \left\langle \mathbf{f}'_{p,k}\left(\mathbf{p}_k, \mathbf{w}^p, \gamma_{1k}\right)\right\rangle, \quad \gamma_{2k} = \frac{(1-\alpha_{1k})\gamma_{1k}}{\alpha_{1k}} \tag{27b}$$

$$\mathbf{v}_k = \frac{1}{1-\alpha_{1k}}\left[\mathbf{f}_{p,k}\left(\mathbf{p}_k, \mathbf{w}^p, \gamma_{1k}\right) - \alpha_{1k}\mathbf{p}_k\right] \tag{27c}$$

$$\mathbf{q}_k = \mathbf{V}'^\top\mathbf{v}_k \tag{27d}$$

$$\alpha_{2k} = \left\langle \mathbf{f}'_{q,k}\left(\mathbf{q}_k, \mathbf{x}_{0,\mathrm{R,I}}, \gamma_{2k}\right)\right\rangle, \quad \gamma_{1,k+1} = \frac{\left(1-\alpha_{2k}\right)\gamma_{2k}}{\alpha_{2k}} \tag{27e}$$

$$\mathbf{u}_{k+1}=\frac{1}{1-\alpha_{2k}}\left[\mathbf{f}_{q,k}\left(\mathbf{q}_k,\mathbf{x}_{0,\mathrm{R,I}},\gamma_{2k}\right)-\alpha_{2k}\mathbf{q}_k\right], \quad (27f)$$

where $\mathbf{V}^{'}=\mathbf{V}^{\top}$ and for $i=1,2$ we have

$$C_i\left(\alpha_i\right)=\frac{1}{1-\alpha_i}, \quad \Gamma_i\left(\gamma_i,\alpha_i\right)=\gamma_i\left[\frac{1}{\alpha_i}-1\right]. \quad (28)$$

Next, we use the algorithm of VAMP-DU and Equations (23)-(26) to derive Recursion (27). Note that (27a) and (27d) can be derived directly. We will derive other equations in Recursion (27). The LMMSE estimator $\mathbf{g}_{1,k}\left(\mathbf{r}_{1k},\gamma_{1k}\right)$ can be written as:

$$\begin{aligned}
&\mathbf{g}_{1,k}\left(\mathbf{r}_{1k},\gamma_{1k}\right)\\
&=\left(\gamma_{wk}\mathbf{A}_{\mathrm{R,I}}^{\top}\mathbf{A}_{\mathrm{R,I}}+\gamma_{1k}\mathbf{I}\right)^{-1}\left(\gamma_{wk}\mathbf{A}_{\mathrm{R,I}}^{\top}\mathbf{A}_{\mathrm{R,I}}\mathbf{x}_{0,\mathrm{R,I}}+\gamma_{wk}\mathbf{A}_{\mathrm{R,I}}^{\top}\mathbf{n}_{\mathrm{R,I}}+\gamma_{1k}\mathbf{r}_{1k}\right)\\
&=\mathbf{x}_{0,\mathrm{R,I}}+\left(\gamma_{wk}\mathbf{A}_{\mathrm{R,I}}^{\top}\mathbf{A}_{\mathrm{R,I}}+\gamma_{1k}\mathbf{I}\right)^{-1}\left(\gamma_{1k}\left(\mathbf{r}_{1k}-\mathbf{x}_{0,\mathrm{R,I}}\right)+\gamma_{wk}\mathbf{A}_{\mathrm{R,I}}^{\top}\mathbf{n}_{\mathrm{R,I}}\right)\\
&=\mathbf{x}_{0,\mathrm{R,I}}+\mathbf{V}\mathbf{f}_{p,k}\left(\mathbf{p}_k,\mathbf{w}^p,\gamma_{1k}\right),
\end{aligned} \quad (29)$$

Then, the divergence can be written as:

$$\begin{aligned}
\alpha_{1k}&=\frac{1}{N}\mathrm{Tr}\left[\frac{\partial\mathbf{g}_{1,k}\left(\mathbf{r}_{1k},\gamma_{1k}\right)}{\partial\mathbf{r}_{1k}}\right]\\
&=\frac{1}{N}\mathrm{Tr}\left[\mathbf{V}^{\top}\mathrm{Diag}\left(\mathbf{f}_{p,k}^{'}\left(\mathbf{p}_k,\mathbf{w}^p,\gamma_{1k}\right)\right)\mathbf{V}\right]\\
&=\left\langle\mathbf{f}_{p,k}^{'}\left(\mathbf{p}_k,\mathbf{w}^p,\gamma_{1k}\right)\right\rangle,
\end{aligned} \quad (30)$$

The line 8 in VAMP-DU algorithm (Appendix A) can be written as:

$$\gamma_{2k}=\eta_{1k}-\gamma_{1k}=\gamma_{1k}\left[\frac{1}{\alpha_{1k}}-1\right]. \quad (31)$$

Equations (30) and (31) derive (27b). Using Equation (27a), we can calculate:

$$\begin{aligned}
\mathbf{q}_k&=\mathbf{r}_{2k}-\mathbf{x}_{0,\mathrm{R,I}}=\left(\eta_{1k}\hat{\mathbf{x}}_{1k}-\gamma_{1k}\mathbf{r}_{1k}\right)/\gamma_{2k}-\mathbf{x}_{0,\mathrm{R,I}}\\
&=\left[\frac{\mathbf{g}_{1,k}(\mathbf{r}_{1,k},\gamma_{1k})-\alpha_{1k}\mathbf{r}_{1,k}}{1-\alpha_{1k}}-\mathbf{x}_{0,\mathrm{R,I}}\right]\\
&=\left[\frac{\mathbf{V}\mathbf{f}_{p,k}\left(\mathbf{p}_k,\mathbf{w}^p,\gamma_{1k}\right)-\alpha_{1k}\mathbf{u}_k}{1-\alpha_{1k}}\right]\\
&=\mathbf{V}\frac{\mathbf{f}_{p,k}\left(\mathbf{p}_k,\mathbf{w}^p,\gamma_{1k}\right)-\alpha_{1k}\mathbf{p}_k}{1-\alpha_{1k}}=\mathbf{V}^{'\top}\frac{\mathbf{f}_{p,k}\left(\mathbf{p}_k,\mathbf{w}^p,\gamma_{1k}\right)-\alpha_{1k}\mathbf{p}_k}{1-\alpha_{1k}},
\end{aligned} \quad (32)$$

which derives (27c). Similarly, we can derive:

$$\alpha_{2k}=\frac{1}{N}\mathrm{Tr}\left[\frac{\partial\mathbf{g}_{2,k}\left(\mathbf{r}_{2k},\gamma_{2k}\right)}{\partial\mathbf{r}_{2k}}\right]=\left\langle\mathbf{f}_{q,k}^{'}\left(\mathbf{q}_k,\mathbf{x}_{0,\mathrm{R,I}},\gamma_{2k}\right)\right\rangle, \quad (33)$$

and

$$\gamma_{1,k+1}=\eta_{2k}-\gamma_{2k}=\gamma_{2k}\left[\frac{1}{\alpha_{2k}}-1\right], \quad (34)$$

which derives (27e). We derive $\mathbf{u}_{k+1}$ as:

$$\begin{aligned}
\mathbf{u}_{k+1}&=\mathbf{r}_{1,k+1}-\mathbf{x}_{0,\mathrm{R,I}}=\left(\eta_{2k}\hat{\mathbf{x}}_{2k}-\gamma_{2k}\mathbf{r}_{2k}\right)/\gamma_{1,k+1}-\mathbf{x}_{0,\mathrm{R,I}}\\
&=\frac{\mathbf{g}_{2,k}(\mathbf{r}_{2,k},\gamma_{2k})-\alpha_{2k}\mathbf{r}_{2,k}}{1-\alpha_{2k}}-\mathbf{x}_{0,\mathrm{R,I}}\\
&=\frac{\mathbf{f}_{q,k}\left(\mathbf{q}_k,\mathbf{x}_{0,\mathrm{R,I}},\gamma_{2k}\right)-\alpha_{2k}\mathbf{q}_k}{1-\alpha_{2k}},
\end{aligned} \quad (35)$$

which derives (27f). Thus, we prove that VAMP-DU is a special case of the Recursion (11). In our CS radar detection problem, we have $\mathbf{q}_{K_{\mathrm{it}}}+\mathbf{x}_{0,\mathrm{R,I}}=\mathbf{w}_{\mathrm{R,I}}+\mathbf{x}_{0,\mathrm{R,I}}=\mathbf{r}_{2,K_{\mathrm{it}}}=\mathbf{r}_{\mathrm{R,I}}$. Therefore, the components $w_{\mathrm{R,I,i}}$ of $\mathbf{w}_{\mathrm{R,I}}=\mathbf{q}_{K_{\mathrm{it}}}$ follow a Gaussian distribution, represented as $w_{\mathrm{R,I},i}\sim\mathcal{N}(0,\sigma_{\mathrm{R,I}}^2)$. Thus, we can derive that the components $w_i$ of $\mathbf{w}=\mathbf{r}-\mathbf{x}_0$ follow a complex Gaussian distribution, denoted as $w_i\sim\mathcal{CN}(0,2\sigma_{\mathrm{R,I}}^2)$. This completes the proof of *Theorem 1*.

□

*Theorem 1* reveals that $w_i$ follows a complex Gaussian distribution $w_i\sim\mathcal{CN}(0,2\sigma_{\mathrm{R,I}}^2)$, which renders the distributions of the test statistic $p(|r_i|\mid H_{1,i})$ and $p(|r_i|\mid H_{0,i})$ analytically derivable. These distributions can be expressed as follows:

$$\begin{aligned}
&p\left(|r_i|\mid H_{0,i}\right)=\frac{|r_i|}{\sigma_{\mathrm{R,I}}^2}\exp\left[-\frac{|r_i|^2}{2\sigma_{\mathrm{R,I}}^2}\right],|r_i|>0,\\
&p\left(|r_i|\mid H_{1,i}\right)=\frac{|r_i|}{\sigma_{\mathrm{R,I}}^2}\exp\left[-\frac{\left(|r_i|^2+|x_{0,i}|^2\right)}{2\sigma_{\mathrm{R,I}}^2}\right]I_0\left(\frac{|x_{0,i}||r_i|}{\sigma_{\mathrm{R,I}}^2}\right),|r_i|>0.
\end{aligned} \quad (36)$$

Here, $x_{0,i}$ is the $i$-th element of $\mathbf{x}_0$ and $I_0(x)$ is the zeroth-order modified Bessel function. To calculate $\sigma^2_{\mathrm{R,I}}$ and implement CS radar detection under a given false alarm rate, we propose PCRD in the following.

## IV. Proposal and Analysis of PCRD

Note that $\mathbf{x}_0$ is not available in practical scenarios, making $\mathbf{w}=\mathbf{r}-\mathbf{x}_0$ unattainable. To calculate $\sigma^2_{\mathrm{R,I}}$ and thus obtain $p(|r_i| \mid H_{1,i})$ and $p(|r_i| \mid H_{0,i})$, which enables the implementation of CS radar detection under a given false alarm rate, in the following subsections, we propose and analyze PCRD based on the property of the sparse solution and pseudo-measurement in VAMP-DU.

### A. PCRD Algorithm

The proposed PCRD algorithm estimates the variance $\sigma^2_{\mathrm{R,I}}$ and the detection threshold $T_{P_{\mathrm{fa}}}$ via iterations, where $T_{P_{\mathrm{fa}}}$ denots the threshold corresponding to the desired false alarm rate $P_{\mathrm{fa}}$. In the $m$-th iteration, estimates $\hat{\mathbf{x}}_{\mathrm{R,I}}{}^{\{m\}}$ and $\hat{\sigma}^2_{\mathrm{R,I}}{}^{\{m\}}$ are obtained, which serve as estimates of $\mathbf{x}_{0,\mathrm{R,I}}$ and $\sigma^2_{\mathrm{R,I}}$, respectively. In the first iteration in PCRD, the sparse solution $\hat{\mathbf{x}}_{\mathrm{R,I}}$ in VAMP-DU can serve as an initial estimate of the true observation scene $\mathbf{x}_{0,\mathrm{R,I}}$, i.e. $\hat{\mathbf{x}}_{\mathrm{R,I}}{}^{\{1\}} = \hat{\mathbf{x}}_{\mathrm{R,I}}$. Note that $\hat{\mathbf{x}}_{\mathrm{R,I}}{}^{\{m\}}$ only serves as an indicator of zero and non-zero entries in $\mathbf{x}_{0,\mathrm{R,I}}$. The numerical values of non-zero elements in $\hat{\mathbf{x}}_{\mathrm{R,I}}{}^{\{m\}}$ and the corresponding elements in $\mathbf{x}_{0,\mathrm{R,I}}$ may exhibit numerical discrepancies. Therefore, for $i \notin \{\mathrm{supp}(\hat{\mathbf{x}}_{\mathrm{R,I}}{}^{\{m\}})\}$, where $x_{0,\mathrm{R,I},i} \approx \hat{x}_{\mathrm{R,I},i}{}^{\{m\}} = 0$, we have:

$$w_{\mathrm{R,I},i} = r_{\mathrm{R,I},i} - x_{0,\mathrm{R,I},i} \approx r_{\mathrm{R,I},i} - \hat{x}_{\mathrm{R,I},i}{}^{\{m\}} = r_{\mathrm{R,I},i}. \quad (37)$$

Thus, the corresponding elements $r_{\mathrm{R,I},i}$ in $\mathbf{r}_{\mathrm{R,I}}$ with $i \notin \{\mathrm{supp}(\hat{\mathbf{x}}_{\mathrm{R,I}}{}^{\{m\}})\}$ can be used as an approximation to estimate the error term elements $w_{\mathrm{R,I},i}, i \notin \{\mathrm{supp}(\hat{\mathbf{x}}_{\mathrm{R,I}}{}^{\{m\}})\}$.

Based on the distribution of the error term $w_{\mathrm{R,I},i} \sim \mathcal{N}(0, \sigma^2_{\mathrm{R,I}})$, by assembling the elements $r_{\mathrm{R,I},i}, i \notin \{\mathrm{supp}(\hat{\mathbf{x}}_{\mathrm{R,I}}{}^{\{m\}})\}$ into a vector $\mathbf{x}_{\mathrm{s,R,I}} \in \mathbb{R}^{L\times 1}$ where $L = 2N - |\mathrm{supp}(\hat{\mathbf{x}}_{\mathrm{R,I}}{}^{\{m\}})|$, the estimate of $\sigma^2_{\mathrm{R,I}}$ in the $m$-th iteration can be expressed as:

$$\hat{\sigma}^2_{\mathrm{R,I}}{}^{\{m\}} = \frac{L}{L-1} \frac{\sum_{i=1}^{L} \left( x_{\mathrm{s,R,I},i} - \frac{\sum_{i=1}^{L} x_{\mathrm{s,R,I},i}}{L} \right)^2}{L}, \quad (38)$$

where $x_{\mathrm{s,R,I},i}$ is the $i$-th element of $\mathbf{x}_{\mathrm{s,R,I}}$. We choose a suitable false alarm rate $P_{\mathrm{fa0}}$ to estimate the observation scene $\mathbf{x}_{0,\mathrm{R,I}}$. According to the Neyman-Pearson criterion and Equation (36), the threshold in the $m$-th iteration in PCRD can be expressed as:

**Algorithm 1** Parameter Convergence Radar Detector

**Require:** $P_{\mathrm{fa}_0}$, $P_{\mathrm{fa}}$, $c_{\mathrm{tol}}$, $\hat{\mathbf{x}}_{\mathrm{R,I}}$, $\mathbf{r}_{\mathrm{R,I}}$, $\mathbf{r}_{\mathrm{R}}$, $\mathbf{r}_{\mathrm{I}}$, $|\mathbf{r}|$
**Initialize** $\hat{\mathbf{x}}_{\mathrm{R,I}}{}^{\{1\}} = \hat{\mathbf{x}}_{\mathrm{R,I}}$, $\hat{\sigma}^2_{\mathrm{R,I}}{}^{\{0\}} = 0$
**for** $m = 1, 2, \ldots, m_{\max}$ **do**
  $L = 2N - \left|\mathrm{supp}\left(\hat{\mathbf{x}}_{\mathrm{R,I}}{}^{\{m\}}\right)\right|$
  $\mathbf{x}_{\mathrm{s,R,I}} = \mathbf{r}_{\mathrm{R,I}}\left[\mathbf{1} - \mathbf{I}\left(\hat{\mathbf{x}}_{\mathrm{R,I}}{}^{\{m\}}\right)\right]$
  $\hat{\sigma}^2_{\mathrm{R,I}}{}^{\{m\}} = \sum_{i=1}^{L}\left(x_{\mathrm{s,R,I},i} - \sum_{i=1}^{L} x_{\mathrm{s,R,I},i}/L\right)^2/(L-1)$
  **if** $\left|\hat{\sigma}^2_{\mathrm{R,I}}{}^{\{m\}} - \hat{\sigma}^2_{\mathrm{R,I}}{}^{\{m-1\}}\right|^2 < c_{\mathrm{tol}}\left|\hat{\sigma}^2_{\mathrm{R,I}}{}^{\{m-1\}}\right|^2$
    $\hat{\sigma}^2_{\mathrm{R,I,PCRD}} = \hat{\sigma}^2_{\mathrm{R,I}}{}^{\{m=m_{\mathrm{cver}}\}}$
    $T_{P_{\mathrm{Pa}}} = \sqrt{-2\hat{\sigma}^2_{\mathrm{R,I,PCRD}} \ln(P_{\mathrm{fa}})}$
    $\hat{\mathbf{x}}_{P_{\mathrm{fa}}} = |\mathbf{r}| \odot \mathbf{I}\left(|\mathbf{r}| > T_{P_{\mathrm{fa}}}\right)$
    **Return** $\hat{\mathbf{x}}_{P_{\mathrm{fa}}}, \hat{\sigma}^2_{\mathrm{R,I,PCRD}}$
  **end if**
  $T_{P_{\mathrm{fa}_0}}{}^{\{m\}} = \sqrt{-2\hat{\sigma}^2_{\mathrm{R,I}}{}^{\{m\}} \ln(P_{\mathrm{fa}_0})}$
  $\hat{\mathbf{x}}_{P_{\mathrm{fa}_0}}{}^{\{m\}} = |\mathbf{r}| \odot \mathbf{I}\left(|\mathbf{r}| > T_{P_{\mathrm{Pa}_0}}{}^{\{m\}}\right)$
  $\hat{\mathbf{x}}_{\mathrm{R,I}}{}^{\{m+1\}} = \left[\mathbf{r}_{\mathrm{R}} \odot \mathbf{I}\left(\hat{\mathbf{x}}_{P_{\mathrm{fa}_0}}{}^{\{m\}} \neq 0\right); \mathbf{r}_{\mathrm{I}} \odot \mathbf{I}\left(\hat{\mathbf{x}}_{P_{\mathrm{fa}_0}}{}^{\{m\}} \neq 0\right)\right]$
**end for**

$$T_{P_{\mathrm{fa0}}}{}^{\{m\}} = \sqrt{-2\hat{\sigma}^2_{\mathrm{R,I}}{}^{\{m\}} \ln\left(P_{\mathrm{fa0}}\right)}. \quad (39)$$

The detection result $\hat{\mathbf{x}}_{P_{\mathrm{fa0}}}{}^{\{m\}}$ corresponding to the false alarm rate $P_{\mathrm{fa0}}$, which serves as an estimate of true observation scene, can be calculated as:

$$\hat{\mathbf{x}}_{P_{\mathrm{fa0}}}{}^{\{m\}} = |\mathbf{r}| \odot \mathbf{I}\left(|\mathbf{r}| > T_{P_{\mathrm{fa0}}}{}^{\{m\}}\right), \quad (40)$$

where $\odot$ represents the Hadamard product. Here, $\mathbf{I}(\mathbf{x})$ is an element-wise indicator function that outputs a binary-value vector of the same size as $\mathbf{x}$. The element of $\mathbf{I}(\mathbf{x})$, can be denoted as $\mathbf{I}(\mathbf{x})_i = 1$ when $x_i \neq 0$ and $\mathbf{I}(\mathbf{x})_i = 0$ when $x_i = 0$. Then, we can obtain $\hat{\mathbf{x}}_{\mathrm{R,I}}{}^{\{m+1\}}$ as:

$$\hat{\mathbf{x}}_{\mathrm{R,I}}{}^{\{m+1\}} = \left[\mathbf{r}_{\mathrm{R}} \odot \mathbf{I}\left(\hat{\mathbf{x}}_{P_{\mathrm{fa0}}}{}^{\{m\}} \neq 0\right); \mathbf{r}_{\mathrm{I}} \odot \mathbf{I}\left(\hat{\mathbf{x}}_{P_{\mathrm{fa0}}}{}^{\{m\}} \neq 0\right)\right], \quad (41)$$

where $\mathbf{r}_{\mathrm{R}} = \mathbf{f}_{\mathrm{R}}(\mathbf{r}_{\mathrm{R,I}}), \mathbf{r}_{\mathrm{I}} = \mathbf{f}_{\mathrm{I}}(\mathbf{r}_{\mathrm{R,I}})$. According to the steps described above, we can obtain $\hat{\sigma}^2_{\mathrm{R,I}}{}^{\{m+1\}}$, which is the estimate of $\sigma^2_{\mathrm{R,I}}$ obtained by PCRD in the $(m+1)$-th iteration. This process is repeated until the estimate $\hat{\sigma}^2_{\mathrm{R,I}}{}^{\{m\}}$ converges, which can be expressed as:

$$\frac{\left|\hat{\sigma}^2_{\mathrm{R,I}}{}^{\{m_{\mathrm{cver}}\}} - \hat{\sigma}^2_{\mathrm{R,I}}{}^{\{m_{\mathrm{cver}}-1\}}\right|}{\hat{\sigma}^2_{\mathrm{R,I}}{}^{\{m_{\mathrm{cver}}-1\}}} < c_{\mathrm{tol}}, \quad (42)$$

where $m_{\mathrm{cver}}$ denotes the PCRD convergent iterations and $c_{\mathrm{tol}} > 0$ is the tolerance level for evaluating convergence. After iterations, we can obtain the estimated error term variance $\hat{\sigma}^2_{\mathrm{R,I,PCRD}} = \hat{\sigma}^2_{\mathrm{R,I}}{}^{\{m_{\mathrm{cver}}\}}$.

Subsequently, the threshold $T_{P_{\mathrm{fa}}}$ corresponding to the desired false alarm rate $P_{\mathrm{fa}}$ can be obtained as:

$$T_{P_{\mathrm{fa}}} = \sqrt{-2\hat{\sigma}^2_{\mathrm{R,I,PCRD}} \ln\left(P_{\mathrm{fa}}\right)}. \tag{43}$$

Consequently, the detection result $\hat{\mathbf{x}}_{P_{\mathrm{fa}}}$ corresponding to the desired false alarm rate $P_{\mathrm{fa}}$ can be calculated as follows:

$$\hat{\mathbf{x}}_{P_{\mathrm{fa}}} = |\mathbf{r}| \odot \mathbf{I}\left(|\mathbf{r}| > T_{P_{\mathrm{fa}}}\right). \tag{44}$$

Detailed steps are provided in Algorithm 1.

*B. Analysis of PCRD*

We analyze the effectiveness of PCRD in terms of variance estimation by the following theorem:

*Theorem 2.* In the proposed PCRD, if $P_{\mathrm{fa}_0} \in (P_{\mathrm{fa,min}}, P_{\mathrm{fa,max}})$, the estimate in each iteration $\hat{\sigma}^2_{\mathrm{R,I}}{}^{\{m\}}$ converges to PCRD estimate $\hat{\sigma}^2_{\mathrm{R,I,PCRD}}$. Specifically, the value of $\hat{\sigma}^2_{\mathrm{R,I,PCRD}}$ is close to the true variance $\sigma^2_{\mathrm{R,I}}$, with their relative squared error (RSE) bounded by:

$$\left|\frac{\hat{\sigma}^2_{\mathrm{R,I,PCRD}} - \sigma^2_{\mathrm{R,I}}}{\sigma^2_{\mathrm{R,I}}}\right|^2 \le P_{\mathrm{fa}_0}^{\frac{2\bar{\sigma}^2_{\mathrm{R,I}}{}^{\{m\}}}{\sigma^2_{\mathrm{R,I}}}} (1 - \ln P_{\mathrm{fa}_0})^2, \tag{45}$$

where $\bar{\sigma}^2_{\mathrm{R,I}}{}^{\{m\}} = \sum_{m=1}^{m_{\mathrm{cver}}} \hat{\sigma}^2_{\mathrm{R,I}}{}^{\{m\}} / m_{\mathrm{cver}}$ denotes the mean of the iterative estimates $\hat{\sigma}^2_{\mathrm{R,I}}{}^{\{m\}}$ over the PCRD convergent iterations $m_{\mathrm{cver}}$. The values of $P_{\mathrm{fa,min}}$ and $P_{\mathrm{fa,max}}$ will be elaborated on in the proof.

*Proof.* We classify the elements of $\mathbf{r}_{\mathrm{R,I}}$ into two categories: $\mathbf{r}_{\mathrm{R,I},H_0}$ and $\mathbf{r}_{\mathrm{R,I},H_1}$. Here, $\mathbf{r}_{\mathrm{R,I},H_0}$ is the vector formed by elements $r_{\mathrm{R,I},i}$ corresponding to $i \notin \{\mathrm{supp}(\mathbf{x}_{0,\mathrm{R,I}})\}$ and $\mathbf{r}_{\mathrm{R,I},H_1}$ is the vector formed by elements $r_{\mathrm{R,I},i}$ corresponding to $i \in \{\mathrm{supp}(\mathbf{x}_{0,\mathrm{R,I}})\}$.

Since the mean squared error (MSE) loss function in VAMP-DU does not impose constraints on sparsity, the solution $\hat{\mathbf{x}}_{\mathrm{R,I}}{}^{\{1\}} = \hat{\mathbf{x}}_{\mathrm{R,I}}$ recovered by VAMP-DU contains more non-zero elements compared to the true values. Specifically, the set $\mathrm{supp}(\hat{\mathbf{x}}_{\mathrm{R,I}}{}^{\{1\}})$ contains more elements than $\mathrm{supp}(\mathbf{x}_{0,\mathrm{R,I}})$ and the value $L = 2N - |\mathrm{supp}(\hat{\mathbf{x}}_{\mathrm{R,I}}{}^{\{1\}})|$ is smaller than $2N - |\mathrm{supp}(\mathbf{x}_{0,\mathrm{R,I}})|$. As a result, $\mathbf{x}_{\mathrm{s,R,I}}$ only includes the smaller values from the vector $\mathbf{r}_{\mathrm{R,I},H_0}$, since elements with larger values are misclassified as false alarm. Thus, the variance of the error term estimated by PCRD in the first iteration $\hat{\sigma}^2_{\mathrm{R,I}}{}^{\{1\}}$ is smaller than $\sigma^2_{\mathrm{R,I}}$, which can be expressed as:

$$\hat{\sigma}^2_{\mathrm{R,I}}{}^{\{1\}} < \sigma^2_{\mathrm{R,I}}. \tag{46}$$

The error term variance estimated in the second iteration $\hat{\sigma}^2_{\mathrm{R,I}}{}^{\{2\}}$ can be expressed as:

$$\hat{\sigma}^2_{\mathrm{R,I}}{}^{\{2\}} = \frac{L}{L-1} \cdot \left( \frac{\sum_{i=1}^{L} x^2_{\mathrm{s,R,I},i}}{L} - \bar{\mathbf{x}}_{\mathrm{s,R,I}}{}^2 \right), \tag{47}$$

where $\bar{\mathbf{x}}_{\mathrm{s,R,I}} = \sum_{i=1}^{L} x_{\mathrm{s,R,I},i} / L$. Based on the Law of Large Numbers, we approximate the mean of $x^2_{\mathrm{s,R,I},i}$ using its expectation $\mathbb{E}(x^2_{\mathrm{s,R,I},i}) = \sum_{i=1}^{L} x^2_{\mathrm{s,R,I},i} / L$. For ease of analysis, a similar strategy as B.Widrow in [37] is employed to interchange the expected values and raw data. Therefore, $\mathbb{E}(\bar{\mathbf{x}}^2_{\mathrm{s,R,I}})$ is utilized as an approximation to $\bar{\mathbf{x}}^2_{\mathrm{s,R,I}}$. Thus, $\hat{\sigma}^2_{\mathrm{R,I}}{}^{\{2\}}$ can be further derived as:

$$\hat{\sigma}^2_{\mathrm{R,I}}{}^{\{2\}} = \frac{L}{L-1}\left[\mathbb{E}\left(x^2_{\mathrm{s,R,I},i}\right) - \mathbb{E}\left(\bar{\mathbf{x}}^2_{\mathrm{s,R,I}}\right)\right] = \mathbb{E}\left(x^2_{\mathrm{s,R,I},i}\right). \tag{48}$$

Based on the Law of Large Numbers, we can derive that:

$$\begin{aligned} \mathbb{E}\left(x^2_{\mathrm{s,R,I},i}\right) &\approx \frac{1}{L}\sum_{i=1}^{L} x^2_{\mathrm{s,R,I},i} = \frac{1}{2}\frac{\sum_{i=1}^{L/2} x^2_{\mathrm{s,R},i} + x^2_{\mathrm{s,I},i}}{\frac{L}{2}} \\ &= \frac{1}{2}\frac{\sum_{i=1}^{L/2} r^2_{\mathrm{s},i}}{\frac{L}{2}} \approx \frac{1}{2}\mathbb{E}\left(r^2_{\mathrm{s},i}\right), \end{aligned} \tag{49}$$

where $x_{\mathrm{s,R},i}$ and $x_{\mathrm{s,I},i}$ are the elements of $\mathbf{x}_{\mathrm{s,R}} = \mathbf{f}_{\mathrm{R}}(\mathbf{x}_{\mathrm{s,R,I}})$ and $\mathbf{x}_{\mathrm{s,I}} = \mathbf{f}_{\mathrm{I}}(\mathbf{x}_{\mathrm{s,R,I}})$, which are the real and imaginary parts of the real-valued formulation $\mathbf{x}_{\mathrm{s,R,I}}$, respectively. Element $r_{\mathrm{s},i}$ is defined as $r_{\mathrm{s},i} = \sqrt{x^2_{\mathrm{s,R},i} + x^2_{\mathrm{s,I},i}}$. Since the elements $x_{\mathrm{s,R},i}$ and $x_{\mathrm{s,I},i}$ satisfy the following inequality:

$$\sqrt{x^2_{\mathrm{s,R},i} + x^2_{\mathrm{s,I},i}} = r_{\mathrm{s},i} < T_{P_{\mathrm{fa}_0}}{}^{\{1\}} = \sqrt{-2\hat{\sigma}^2_{\mathrm{R,I}}{}^{\{1\}} \ln\left(P_{\mathrm{fa}_0}\right)}, \tag{50}$$

we can further derive $\mathbb{E}\left(r^2_{\mathrm{s},i}\right)/2$ as:

$$\frac{1}{2}\mathbb{E}\left(r^2_{\mathrm{s},i}\right) = \frac{1}{2} n_{\mathrm{e}} \int_0^{T_{P_{\mathrm{fa}0}}{}^{\{1\}}} r^2 \frac{r}{\sigma^2_{\mathrm{R,I}}} \exp\left[-\frac{r^2}{2\sigma^2_{\mathrm{R,I}}}\right] dr, \tag{51}$$

where the normalization factor $n_{\mathrm{e}}$ can be expressed as:

$$n_{\mathrm{e}} = \frac{1}{\int_0^{T_{P_{\mathrm{fa}0}}{}^{\{1\}}} \frac{r}{\sigma^2_{\mathrm{R,I}}} \exp\left[-\frac{r^2}{2\sigma^2_{\mathrm{R,I}}}\right] dr}. \tag{52}$$

We choose a small value for $P_{\mathrm{fa}_0}$ in PCRD, thus the value for $(T_{P_{\mathrm{fa}_0}}{}^{\{1\}})^2$ is relatively large and the effect of $n_{\mathrm{e}}$ on the calculation is minimal. We can approximate $n_{\mathrm{e}} = 1$ and $\hat{\sigma}^2_{\mathrm{R,I}}{}^{\{2\}}$ can be further derived as:

$$\hat{\sigma}^2_{\mathrm{R,I}}{}^{\{2\}} = \frac{1}{2}\mathbb{E}\left(r^2_{\mathrm{s},i}\right) = \sigma^2_{\mathrm{R,I}} - \frac{1}{2}\exp\left(-\frac{(T_{P_{\mathrm{fa}_0}}{}^{\{1\}})^2}{2\sigma^2_{\mathrm{R,I}}}\right)\left(2\sigma^2_{\mathrm{R,I}} + (T_{P_{\mathrm{fa}_0}}{}^{\{1\}})^2\right). \tag{53}$$

Similarly, by continuing the derivation in this manner, we can obtain:

$$\hat{\sigma}_{\mathrm{R,I}}^{2}{}^{\{m+1\}} = \sigma_{\mathrm{R,I}}^{2} - \frac{1}{2}\exp\left(-\frac{(T_{P_{\mathrm{fa}_0}}{}^{\{m\}})^2}{2\sigma_{\mathrm{R,I}}^{2}}\right)\left(2\sigma_{\mathrm{R,I}}^{2} + (T_{P_{\mathrm{fa}_0}}{}^{\{m\}})^2\right). \tag{54}$$

For the convenience of further analysis, we define the function $f(T)$ as:

$$f(T) = \sigma_{\mathrm{R,I}}^{2} - \frac{1}{2}\exp\left(-\frac{T^2}{2\sigma_{\mathrm{R,I}}^{2}}\right)\left(2\sigma_{\mathrm{R,I}}^{2} + T^2\right),\ T > 0. \tag{55}$$

Since $f(T)$ is monotonically increasing with respect to $T$. Let $P_{\mathrm{fa,max1}} > 0$ satisfy $\hat{\sigma}_{\mathrm{R,I}}^{2}{}^{\{1\}} = f\left(T_{P_{\mathrm{fa,max1}}}{}^{\{1\}}\right)$ .When $0 < P_{\mathrm{fa}_0} < P_{\mathrm{fa,max1}}$, the following equation holds:

$$\hat{\sigma}_{\mathrm{R,I}}^{2}{}^{\{2\}} = f\left(T_{P_{\mathrm{fa}_0}}{}^{\{1\}}\right) > f\left(T_{P_{\mathrm{fa,max1}}}{}^{\{1\}}\right) = \hat{\sigma}_{\mathrm{R,I}}^{2}{}^{\{1\}}. \tag{56}$$

Thus, we can obtain:

$$\hat{\sigma}_{\mathrm{R,I}}^{2}{}^{\{m+1\}} = f\left(T_{P_{\mathrm{fa}_0}}{}^{\{m\}}\right) > f\left(T_{P_{\mathrm{fa}_0}}{}^{\{m-1\}}\right) = \hat{\sigma}_{\mathrm{R,I}}^{2}{}^{\{m\}}. \tag{57}$$

By substituting Equation (39) into Equation (54), we can obtain:

$$\hat{\sigma}_{\mathrm{R,I}}^{2}{}^{\{m+1\}} = g\left(\hat{\sigma}_{\mathrm{R,I}}^{2}{}^{\{m\}}\right), \tag{58}$$

where function $g(\sigma^2)$ is defined as:

$$g\left(\sigma^2\right) = \sigma_{\mathrm{R,I}}^{2} - \frac{1}{2}\exp\left(\frac{\sigma^2 \ln\left(P_{\mathrm{fa}_0}\right)}{\sigma_{\mathrm{R,I}}^{2}}\right)\left(2\sigma_{\mathrm{R,I}}^{2} - 2\sigma^2 \ln\left(P_{\mathrm{fa}_0}\right)\right). \tag{59}$$

Let $P_{\mathrm{fa,max2}} \in (0,\ \exp(-2\sigma_{\mathrm{R,I}}^{2} / \hat{\sigma}_{\mathrm{R,I}}^{2}{}^{\{1\}})\ ]$ satisfy (if there exists, or $P_{\mathrm{fa,max2}} = \exp(-2\sigma_{\mathrm{R,I}}^{2} / \hat{\sigma}_{\mathrm{R,I}}^{2}{}^{\{1\}})$ ):

$$P_{\mathrm{fa,max2}}^{\frac{\hat{\sigma}_{\mathrm{R,I}}^{2}{}^{\{1\}}}{\sigma_{\mathrm{R,I}}^{2}}} \ln^2\left(P_{\mathrm{fa,max2}}\right) = 1. \tag{60}$$

When $0 < P_{\mathrm{fa}_0} < P_{\mathrm{fa,max2}}$, we can obtain:

$$0 < g'\left(\hat{\sigma}_{\mathrm{R,I}}^{2}{}^{\{m\}}\right) < P_{\mathrm{fa}_0}^{\frac{\hat{\sigma}_{\mathrm{R,I}}^{2}{}^{\{1\}}}{\sigma_{\mathrm{R,I}}^{2}}} \ln^2\left(P_{\mathrm{fa}_0}\right) < 1. \tag{61}$$

According to the Lagrange Mean Value Theorem [38], for any $\sigma_1^2, \sigma_2^2 \in [\hat{\sigma}_{\mathrm{R,I}}^{2}{}^{\{1\}},\ \sigma_{\mathrm{R,I}}^{2}]$, there exists $\sigma_0^2 \in (\sigma_1^2,\ \sigma_2^2)$ such that the following equation holds:

$$\begin{aligned}\frac{g\left(\sigma_2^2\right) - g\left(\sigma_1^2\right)}{\sigma_2^2 - \sigma_1^2} &= \frac{\left|g\left(\sigma_2^2\right) - g\left(\sigma_1^2\right)\right|}{\left|\sigma_2^2 - \sigma_1^2\right|} \\ &= g'\left(\sigma_0^2\right) < P_{\mathrm{fa}_0}^{\frac{\hat{\sigma}_{\mathrm{R,I}}^{2}{}^{\{1\}}}{\sigma_{\mathrm{R,I}}^{2}}} \ln^2\left(P_{\mathrm{fa}_0}\right) < 1.\end{aligned} \tag{62}$$

According to the Banach Fixed-Point Theorem [39], on the interval $[\hat{\sigma}_{\mathrm{R,I}}^{2}{}^{\{1\}},\ \sigma_{\mathrm{R,I}}^{2}]$, the iteration $\hat{\sigma}_{\mathrm{R,I}}^{2}{}^{\{m+1\}} = g(\hat{\sigma}_{\mathrm{R,I}}^{2}{}^{\{m\}})$ has a unique convergence fixed point. The fixed point is the error term variance estimated by the PCRD $\hat{\sigma}_{\mathrm{R,I,PCRD}}^{2}$, which can be represented as:

$$\hat{\sigma}_{\mathrm{R,I,PCRD}}^{2} = g\left(\hat{\sigma}_{\mathrm{R,I,PCRD}}^{2}\right). \tag{63}$$

By substituting Equation (59) into Equation (63), we can obtain:

$$\left|\frac{\hat{\sigma}_{\mathrm{R,I,PCRD}}^{2} - \sigma_{\mathrm{R,I}}^{2}}{\sigma_{\mathrm{R,I}}^{2}}\right|^2 \le P_{\mathrm{fa}_0}^{\frac{2\hat{\sigma}_{\mathrm{R,I}}^{2}{}^{\{m\}}}{\sigma_{\mathrm{R,I}}^{2}}} (1 - \ln P_{\mathrm{fa}_0})^2. \tag{64}$$

According to the Banach Fixed-Point Theorem [39], the distance between the current value $\hat{\sigma}_{\mathrm{R,I}}^{2}{}^{\{m\}}$ and the fixed point $\hat{\sigma}_{\mathrm{R,I,PCRD}}^{2}$ decreases exponentially with each iteration.

Note that the above derivation assumes that the elements of $\mathbf{x}_{\mathrm{s,R,I}}$ are all from $\mathbf{r}_{\mathrm{R,I},H_0}$ and do not contain the elements from $\mathbf{r}_{\mathrm{R,I},H_1}$. Let $P_{\mathrm{fa,min}}$ satisfy:

$$\sqrt{-2\hat{\sigma}_{\mathrm{R,I,PCRD}}^{2} \ln\left(P_{\mathrm{fa,min}}\right)} = \min_i\left(\left|r_{\mathrm{R},H_1,i} + j r_{\mathrm{I},H_1,i}\right|\right), \tag{65}$$

where $r_{\mathrm{R},H_1,i}$ and $r_{\mathrm{I},H_1,i}$ are the elements of $\mathbf{r}_{\mathrm{R},H_1} = \mathbf{f}_{\mathrm{R}}(\mathbf{r}_{\mathrm{R,I},H_1})$ and $\mathbf{r}_{\mathrm{I},H_1} = \mathbf{f}_{\mathrm{I}}(\mathbf{r}_{\mathrm{R,I},H_1})$, which are the real and imaginary parts of the real-valued formulation $\mathbf{r}_{\mathrm{R,I},H_1}$, respectively. Thus, we can conclude that if $P_{\mathrm{fa,min}} < P_{\mathrm{fa}_0} < P_{\mathrm{fa,max}} = \min(P_{\mathrm{fa,max1}}, P_{\mathrm{fa,max2}})$, the PCRD estimate $\hat{\sigma}_{\mathrm{R,I,PCRD}}^{2}$ is close to the true variance $\sigma_{\mathrm{R,I}}^{2}$ of the error term, with their RSE bounded by Equation (45). This completes the proof of *Theorem 2.*

□

Note that it is not possible to obtain exact values for $P_{\mathrm{fa,min}}$, $P_{\mathrm{fa,max1}}$ and $P_{\mathrm{fa,max2}}$ in the experiment, since we do not know the values of $\sigma_{\mathrm{R,I}}^{2}$ and $\mathbf{r}_{\mathrm{R,I},H_1}$ previously. We can select a value for $P_{\mathrm{fa}_0}$ based on empirical observations.

*Theorem 2* reveals that the PCRD can accurately estimate the variance of the error term $\sigma_{\mathrm{R,I}}^{2}$, an unknown parameter of the test statistic distributions $p(|r_i| \mid H_{1,i})$ and $p(|r_i| \mid H_{0,i})$, by leveraging the property of sparse solution and pseudo-measurement in VAMP-DU. Consequently, using accurate distributions of the test statistic $p(|r_i| \mid H_{1,i})$ and $p(|r_i| \mid H_{0,i})$, PCRD can accurately set the detection threshold under a given false alarm rate, as expressed in Equation (43).

## V. Simulation Results

### *A. The Gaussianity of Error Term*

In this subsection, we validate the Gaussianity of the error term $\mathbf{w}$. The real part of the error term $\mathbf{w}_{\mathrm{R}} = \mathrm{Re}(\mathbf{w}) \in \mathbb{R}^{N\times 1}$ can be divided into $\mathbf{w}_{\mathrm{R,Tar}} \in \mathbb{R}^{L_0\times 1}$ corresponding to the positions where there are targets present and $\mathbf{w}_{\mathrm{R,NoTar}} \in \mathbb{R}^{(N-L_0)\times 1}$ where there are no targets present. Similarly, the imaginary part of the error term $\mathbf{w}_{\mathrm{I}} = \mathrm{Im}(\mathbf{w}) \in \mathbb{R}^{N\times 1}$ can be divided into $\mathbf{w}_{\mathrm{I,Tar}} \in \mathbb{R}^{L_0\times 1}$ and

TABLE I

JARQUE-BERA TEST FOR REAL AND IMAGINARY PARTS OF ERROR TERM $\mathbf{w}$ CORRESPONDING TO POSITIONS WITH AND WITHOUT TARGETS: $\mathbf{w}_{\mathrm{R,Tar}}$, $\mathbf{w}_{\mathrm{R,NoTar}}$, $\mathbf{w}_{\mathrm{I,Tar}}$ AND $\mathbf{w}_{\mathrm{I,NoTar}}$

| Jarque-Bera Test Metrics | Real Part (Positions with Target) $\mathbf{w}_{\mathrm{R,Tar}}$ | Real Part (Positions without Target) $\mathbf{w}_{\mathrm{R,NoTar}}$ | Imaginary Part (Positions with Target) $\mathbf{w}_{\mathrm{I,Tar}}$ | Imaginary Part (Positions without Target) $\mathbf{w}_{\mathrm{I,NoTar}}$ |
|---|---|---|---|---|
| p-values | 0.3722 | 0.3764 | 0.3743 | 0.3719 |

Note: This table validates the Gaussianity of the error term, since p-values are much larger than the significance level $\alpha = 0.05$, indicating that the null hypothesis of Gaussianity cannot be rejected.

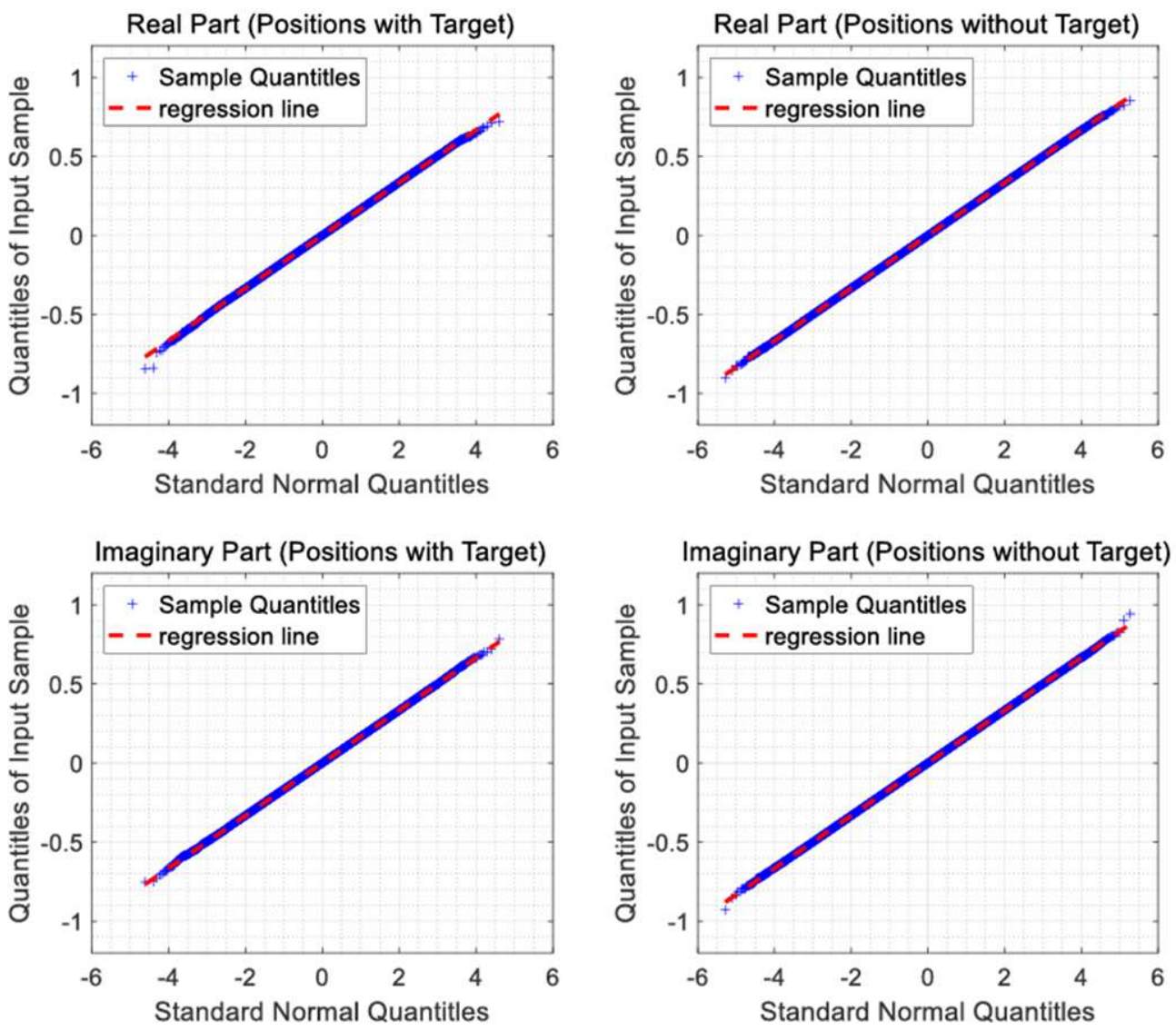

Fig. 1. QQ plots of real and imaginary parts of error term $\mathbf{w}$ corresponding to positions with and without targets: $\mathbf{w}_{\mathrm{R,Tar}}$, $\mathbf{w}_{\mathrm{R,NoTar}}$, $\mathbf{w}_{\mathrm{I,Tar}}$ and $\mathbf{w}_{\mathrm{I,NoTar}}$. This figure validates that the components of $\mathbf{w}_{\mathrm{R,Tar}}$, $\mathbf{w}_{\mathrm{R,NoTar}}$, $\mathbf{w}_{\mathrm{I,Tar}}$ and $\mathbf{w}_{\mathrm{I,NoTar}}$ follow a Gaussian distribution.

$\mathbf{w}_{\mathrm{I,NoTar}} \in \mathbb{R}^{(N-L_0)\times 1}$ corresponding to the positions with and without targets, respectively. To validate the Gaussianity of the real and imaginary parts of the error term corresponding to positions with and without targets: $\mathbf{w}_{\mathrm{R,Tar}}$, $\mathbf{w}_{\mathrm{R,NoTar}}$, $\mathbf{w}_{\mathrm{I,Tar}}$, $\mathbf{w}_{\mathrm{I,NoTar}}$, we employ quantile-quantile (QQ) plots and Jarque-Bera tests [40], subsequently.

In the simulations, the training and testing data parameter set is denoted as $\boldsymbol{\varphi} = \{a_{\min}, a_{\max}, -\pi, \pi, \rho_{\min}, \rho_{\max}, \mathrm{SNR}_{\min}, \mathrm{SNR}_{\max}\}$. The generated training observation scene $\mathbf{x}_{0,\mathrm{R,I,Train}}$ can be expressed as:

$$\mathbf{x}_{0,\mathrm{R,I,Train}} = \mathbf{a} \odot \exp(j\boldsymbol{\Phi}) \odot \mathbf{Q}, \tag{66}$$

where the elements of $\mathbf{a} \in \mathbb{R}^{N\times 1}$ are uniformly distributed within the interval $[a_{\min}, a_{\max}]$. Each element $\Phi_i$ of $\boldsymbol{\Phi} \in \mathbb{R}^{N\times 1}$ is uniformly distributed within the interval $[-\pi, \pi]$. Each element $Q_i$ of $\mathbf{Q} \in \mathbb{R}^{N\times 1}$ is with a PDF $f_Q(q)$, which can be represented as:

$$Q_i \sim f_Q(q) = (1-\rho)\delta(q) + \rho\delta(q-1), \tag{67}$$

where $\delta(\cdot)$ is the Dirac delta function. The signal density $\rho$ is uniformly distributed within the interval $[\rho_{\min}, \rho_{\max}]$.

We simulate a partial observation scenario for radar systems employing stepped-frequency waveforms. In this case, we set the observation matrix as partial Fourier matrix with the dimension as $M = 600, N = 1000$. The parameters for the testing data are set as $\rho_{\min} = \rho_{\max} = 0.03$, $\mathrm{SNR}_{\min} = \mathrm{SNR}_{\max} = 13$ dB, $a_{\min} = a_{\max} = 1.0$. The number of layers for the VAMP-DU is set to $K_{\mathrm{it}} = 7$. The training data parameters for VAMP-DU are set as $\rho_{\min} = 0.01$, $\rho_{\max} = 0.05$, $\mathrm{SNR}_{\min} = 8$ dB, $\mathrm{SNR}_{\max} = 18$ dB, $a_{\min} = 0.7$, $a_{\max} = 1.3$. We conducted 10,000 Monte Carlo experiments to obtain the results in Fig. 1 and Table I.

Fig. 1 shows the QQ plots comparing the real and imaginary parts of error term $\mathbf{w}$ corresponding to positions with and without targets: $\mathbf{w}_{\mathrm{R,Tar}}$, $\mathbf{w}_{\mathrm{R,NoTar}}$, $\mathbf{w}_{\mathrm{I,Tar}}$ and $\mathbf{w}_{\mathrm{I,NoTar}}$, with the standard normal distribution. The quantiles of $\mathbf{w}_{\mathrm{R,Tar}}$, $\mathbf{w}_{\mathrm{R,NoTar}}$, $\mathbf{w}_{\mathrm{I,Tar}}$ and $\mathbf{w}_{\mathrm{I,NoTar}}$ compared to the standard normal distribution fall on the dashed line, confirming that the components of $\mathbf{w}_{\mathrm{R,Tar}}$, $\mathbf{w}_{\mathrm{R,NoTar}}$, $\mathbf{w}_{\mathrm{I,Tar}}$ and $\mathbf{w}_{\mathrm{I,NoTar}}$ are Gaussian-distributed.

Table I shows the Jarque-Bera tests employed on real and imaginary parts of error term $\mathbf{w}$ corresponding to positions with and without targets: $\mathbf{w}_{\mathrm{R,Tar}}$, $\mathbf{w}_{\mathrm{R,NoTar}}$, $\mathbf{w}_{\mathrm{I,Tar}}$ and $\mathbf{w}_{\mathrm{I,NoTar}}$. The Jarque-Bera test validates whether the input variable exhibits the skewness and kurtosis characteristic of a Gaussian distribution. In Jarque-Bera test, a larger output p-value indicates stronger evidence that the sample data follows a Gaussian distribution. Table I demonstrates that the p-values for $\mathbf{w}_{\mathrm{R,Tar}}$, $\mathbf{w}_{\mathrm{R,NoTar}}$, $\mathbf{w}_{\mathrm{I,Tar}}$ and $\mathbf{w}_{\mathrm{I,NoTar}}$ are much higher than the significance level $\alpha = 0.05$, indicating that the components of $\mathbf{w}_{\mathrm{R,Tar}}$, $\mathbf{w}_{\mathrm{R,NoTar}}$, $\mathbf{w}_{\mathrm{I,Tar}}$ and $\mathbf{w}_{\mathrm{I,NoTar}}$ follow a Gaussian distribution.

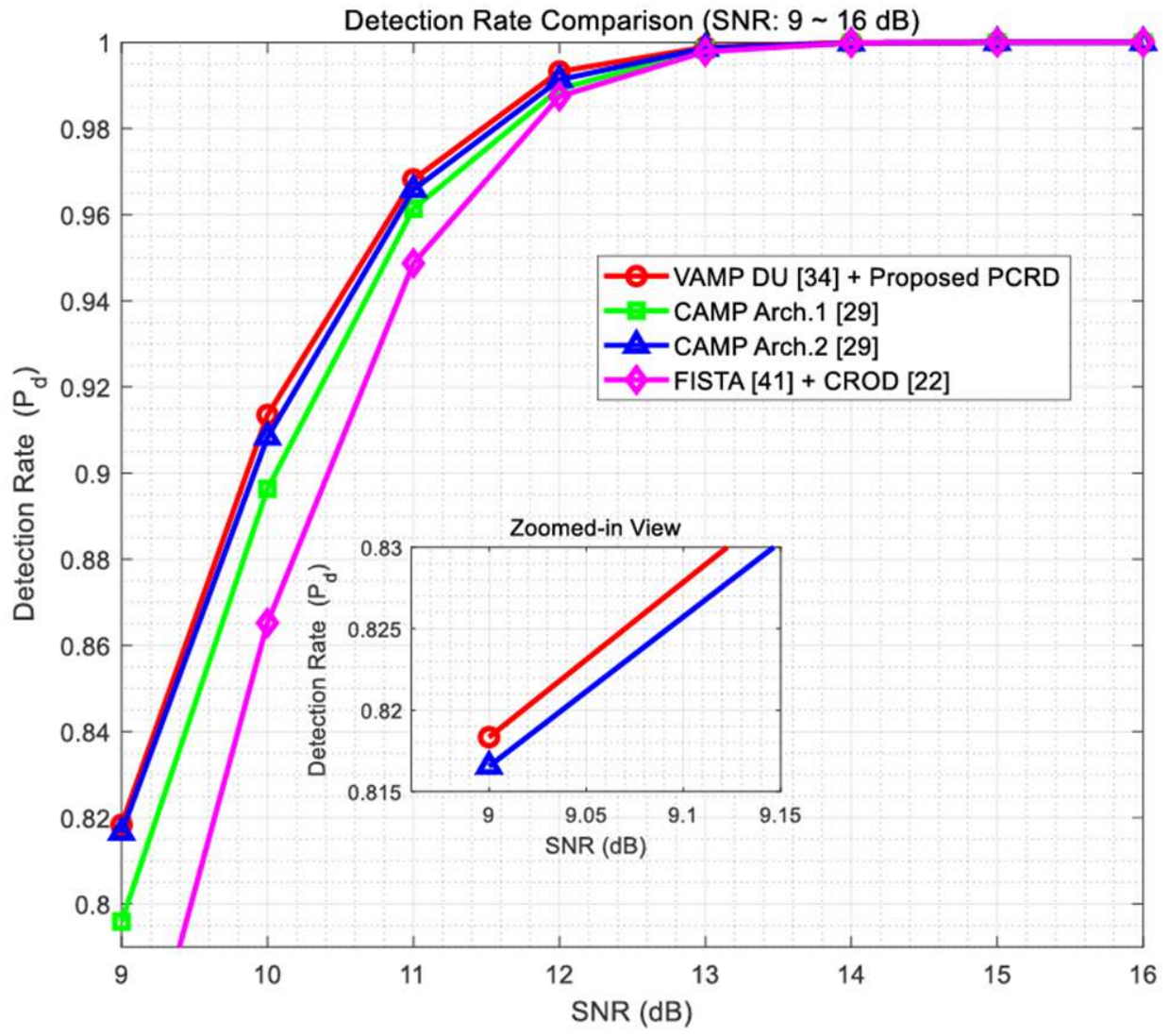

(a)

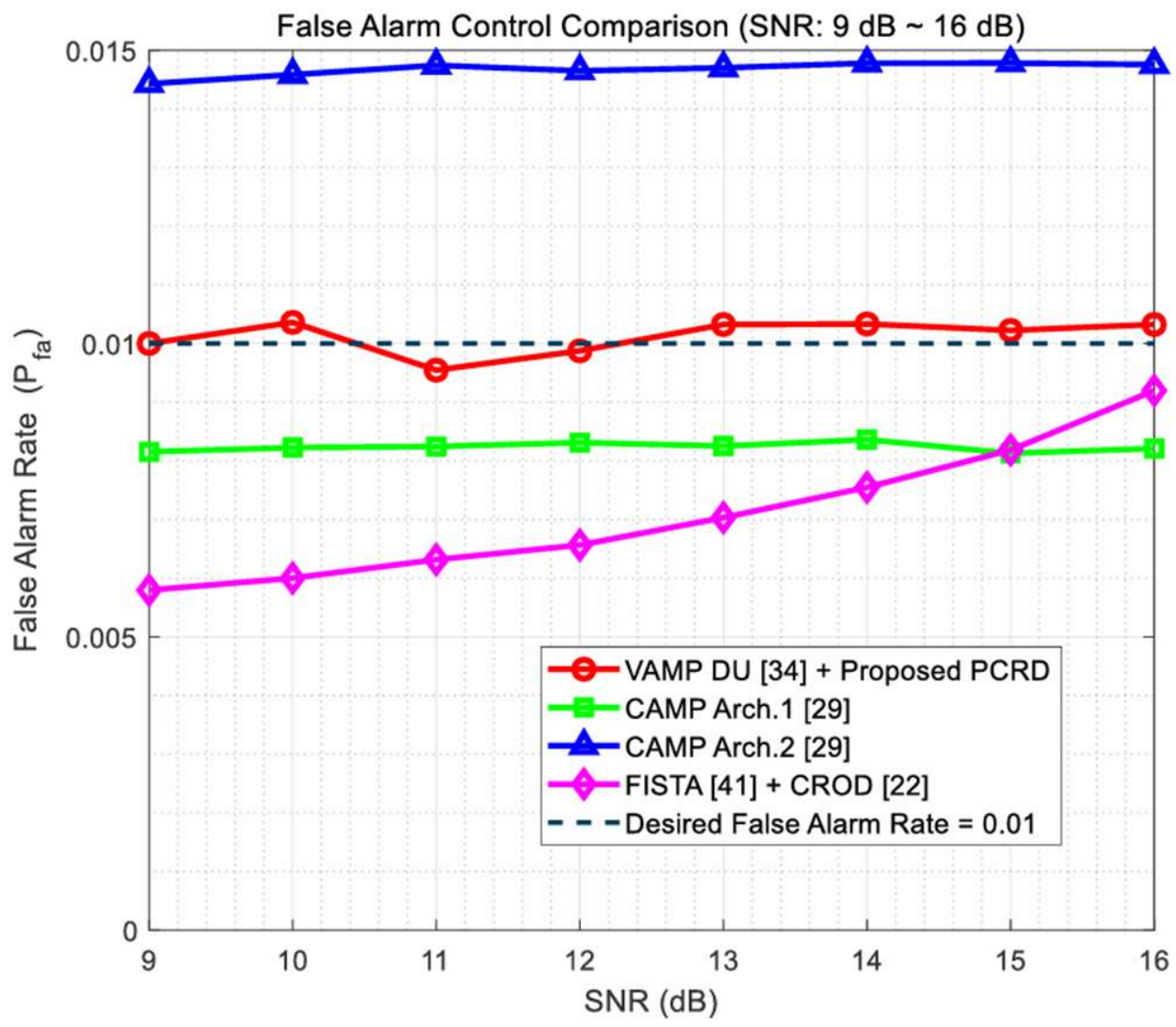

(b)

Fig. 2. Performance comparison of different detection schemes with desired false alarm rate = 0.01, SNR from 9 to 16 dB (a) shows that VAMP-DU combined with PCRD achieves better target detection performance; (b) shows that VAMP-DU combined with PCRD exhibits superior false alarm control capability.

## B. The Performance of PCRD

In the numerical experiments, we set the dimensions of the partial Fourier observation matrix as $M = 600$ , $N = 1000$ . The parameters for the testing data are set as $\rho_{\min} = \rho_{\max} = 0.03$ , $a_{\min} = a_{\max} = 1.0$ . The SNR for testing data ranges from 9 dB to 16 dB. The number of layers for the VAMP-DU is set to $K_{\text{it}} = 7$ . The training data parameters for VAMP-DU are set as $\rho_{\min} = 0.01$ , $\rho_{\max} = 0.05$ , $\text{SNR}_{\min} = 8$ dB , $\text{SNR}_{\max} = 18$ dB , $a_{\min} = 0.7$ , $a_{\max} = 1.3$ . We conducted 2,000 Monte Carlo experiments to obtain Figs. 2 and 3.

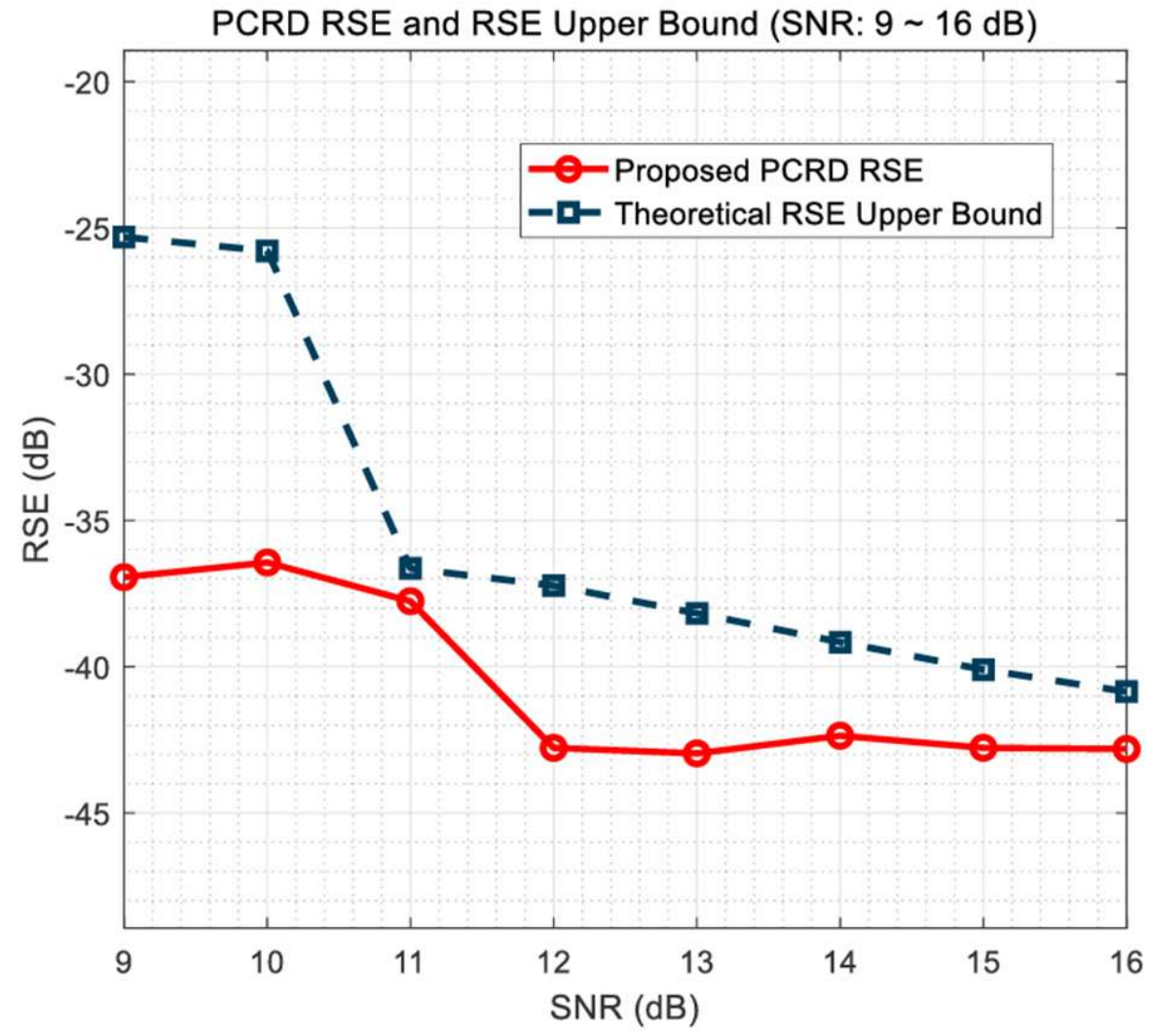

Fig. 3. RSE between PCRD estimate $\hat{\sigma}^2_{\text{R,I,PCRD}}$ and true variance $\sigma^2_{\text{R,I}}$. This figure shows that the low RSE curve lies below the upper bound derived from *Theorem 2*, validating *Theorem 2* and the accuracy of PCRD.

When the probability of false alarm is set to 0.01, Fig. 2 compares the target detection performance and the false alarm control performance, achieved by the VAMP-DU-based PCRD with that of other CS radar detection schemes, including two architectural frameworks of CAMP, FISTA [41] sparse recovery algorithm combined with CROD (CROD is with precise knowledge of the AWGN power $\sigma^2$ while other algorithms operate without such pre-acquired information). When SNR is below 10 dB, the $P_{\text{fa}_0} = 4\times10^{-3}$ in PCRD and when SNR is above 10 dB, the $P_{\text{fa}_0} = 1\times10^{-3}$ in PCRD. In Fig. 2, we demonstrate that VAMP-DU combined with PCRD can achieve the highest target detection rate and the most accurate false alarm control, superior to the performance of two architectural frameworks of CAMP and FISTA combined with CROD.

In Fig. 3, we demonstrate that the value of PCRD estimate $\hat{\sigma}^2_{\text{R,I,PCRD}}$ is close to the true variance $\sigma^2_{\text{R,I}}$, with RSE bounded above by the theoretical upper bound derived from *Theorem 2*, which validates the accuracy of PCRD.

Next, a typical value SNR = 13 dB is selected from the SNR range with $P_{\text{fa}_0} = 1\times10^{-3}$ in PCRD to elaborate on the performance of the VAMP-DU-based PCRD, with further analysis presented in the following.

In the following numerical experiments, we set the dimensions of the partial Fourier observation matrix as $M = 200$ , $N = 256$ . The parameters for the testing data are set as $\rho_{\min} = \rho_{\max} = 0.02$ , $\text{SNR} = 13$ dB , $a_{\min} = a_{\max} = 1.0$ . The number of layers for the VAMP-DU is set to $K_{\text{it}} = 7$ . The training data parameters for VAMP-DU are set as $\rho_{\min} = 0.01$ , $\rho_{\max} = 0.03$ , $\text{SNR}_{\min} = 8$ dB , $\text{SNR}_{\max} = 18$ dB , $a_{\min} = 0.7$ , $a_{\max} = 1.3$ . We conducted 10,000 Monte Carlo experiments.

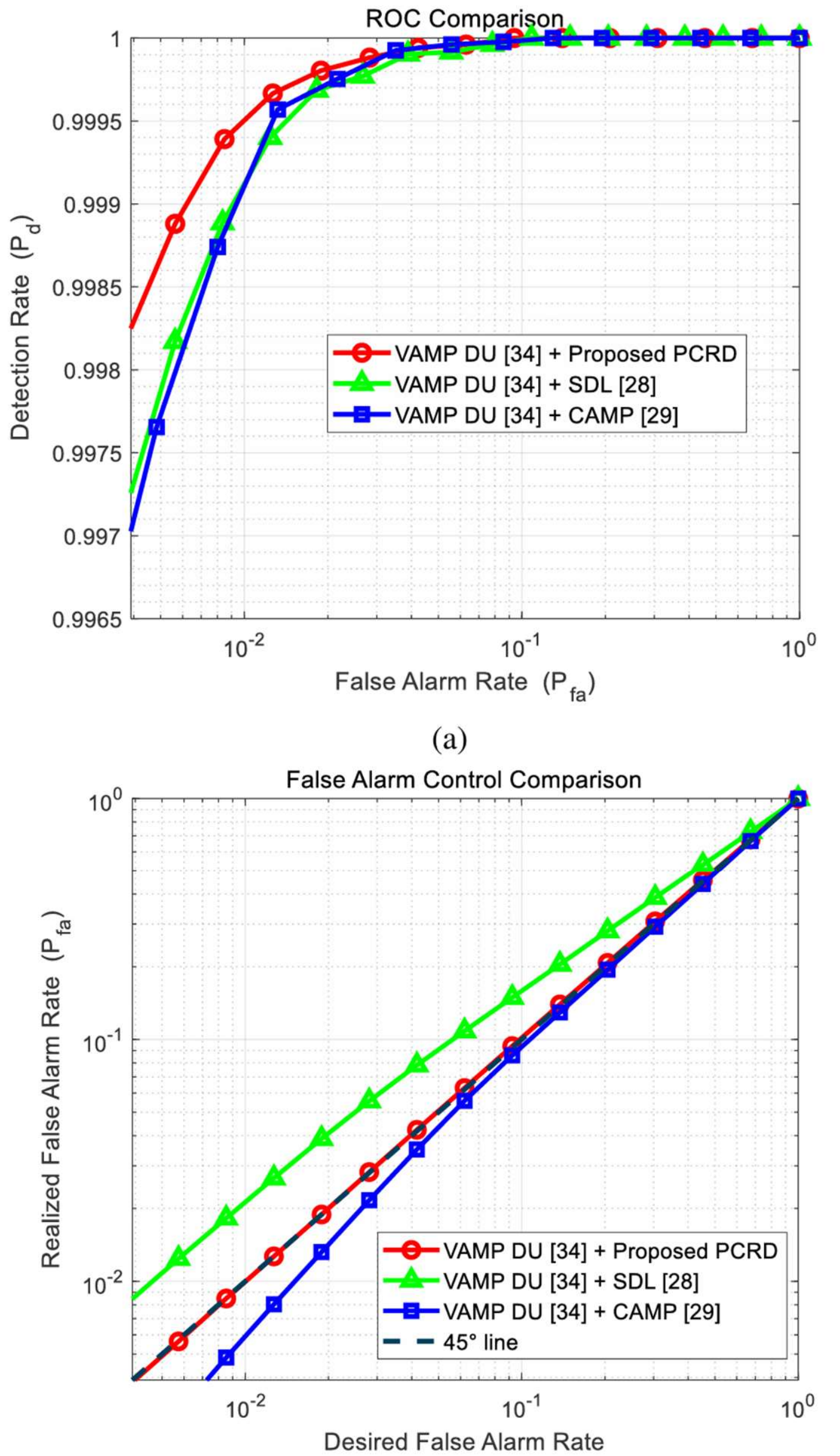

Fig. 4. Performance comparison of VAMP-DU combined with PCRD, CAMP and SDL, at SNR = 13 dB: (a) shows that PCRD achieves better target detection performance; (b) shows that PCRD exhibits superior false alarm control capability.

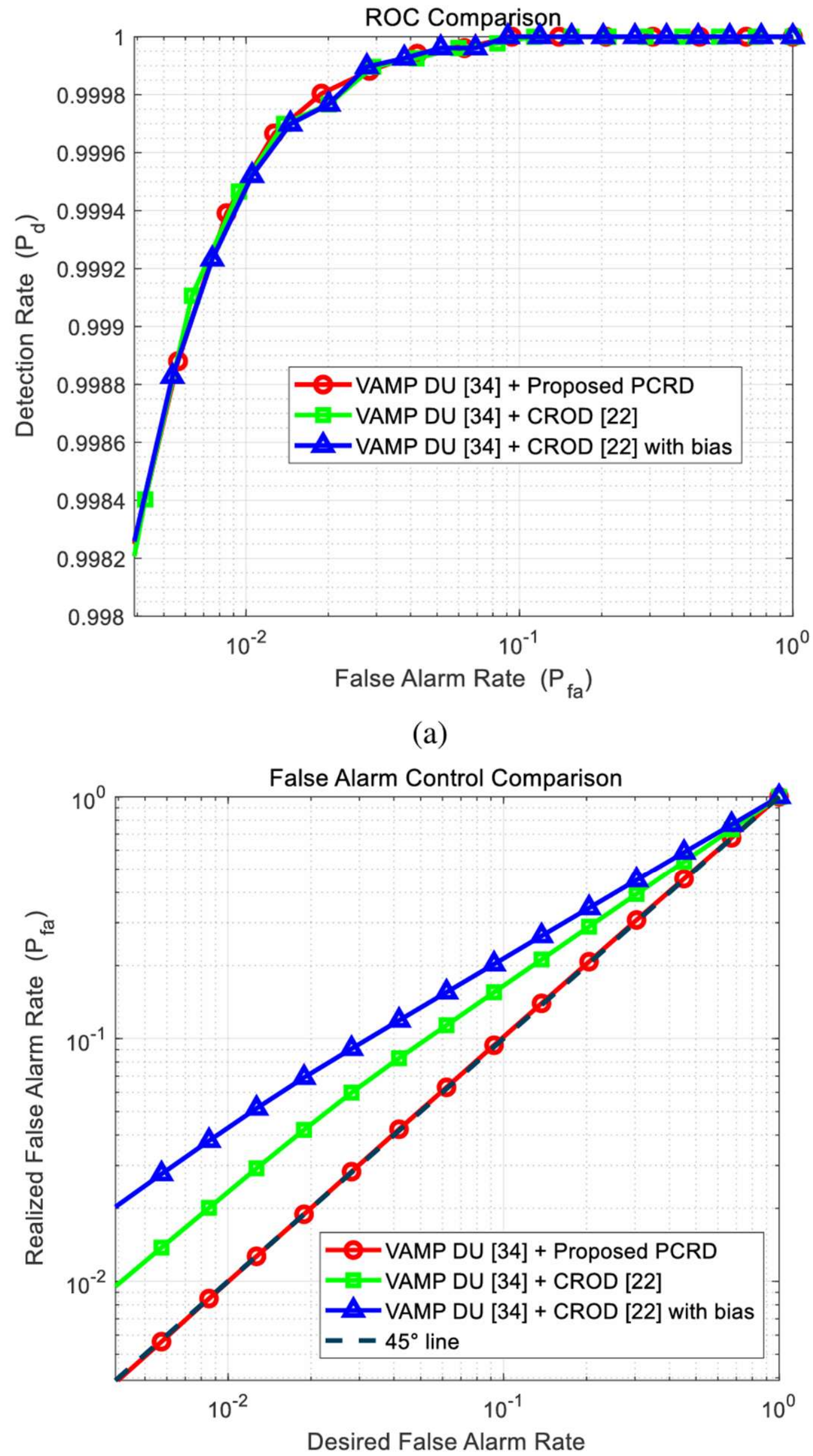

Fig. 5. Performance comparison of VAMP-DU combined with PCRD and CROD, at SNR = 13 dB: (a) shows that PCRD requires no AWGN power knowledge as CROD does, yet has comparable detection performance; (b) shows that PCRD exhibits superior false alarm control capability.

Fig. 4 compares the target detection and false alarm control performance of VAMP-DU combined with PCRD, CAMP and SDL [28], which are the detection algorithms that do not require prior knowledge of the AWGN power. Here, CAMP and SDL use the detection frameworks in [22]. As shown in Fig. 4 (a), the detection performance of VAMP-DU combined with PCRD outperforms that of VAMP-DU combined with CAMP and SDL.

In Fig. 4 (b), the horizontal axis represents the desired false alarm rate, and the vertical axis represents the false alarm rate achieved. The closer the false alarm rate realized to the desired false alarm rate, the more accurate the threshold is set under the given false alarm rate. The $45^{\circ}$ line serves as a reference line where the desired false alarm rate is precisely equal to the achieved one, enabling intuitive comparison between different methods. As shown in Fig. 4 (b), the accuracy of false alarm control of VAMP-DU combined with PCRD is superior to that of VAMP-DU combined with CAMP and SDL.

Fig. 5 compares the target detection and false alarm control performance of VAMP-DU combined with PCRD, CROD, and the scenario in which CROD retrieves AWGN power with a bias set to $\sigma_{\text{bias}} = 0.9\sigma$. From Fig. 5 (a), we can find that the detection performance of VAMP-DU combined with CROD is similar to that when combined with PCRD. However, unlike

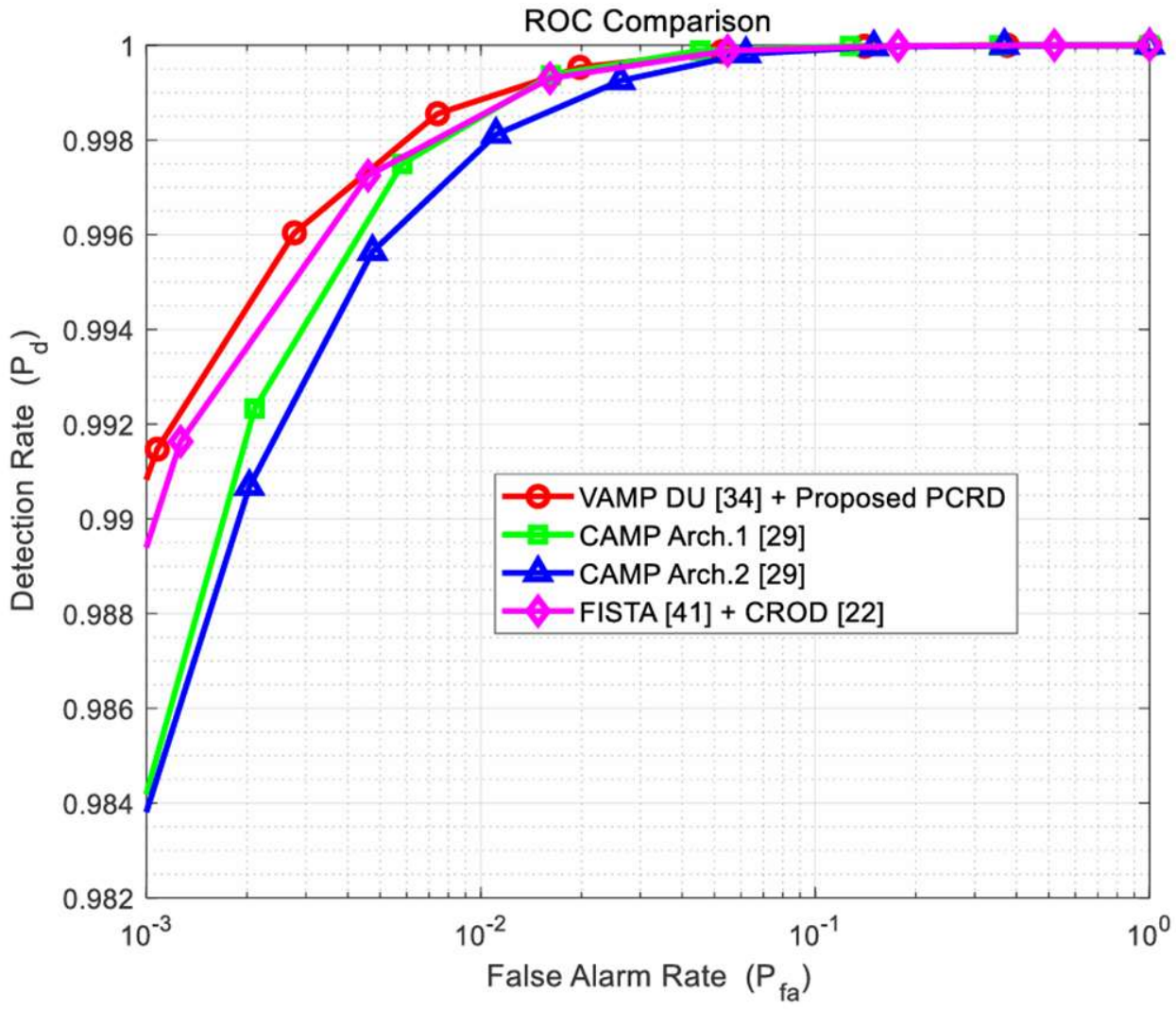

(a)

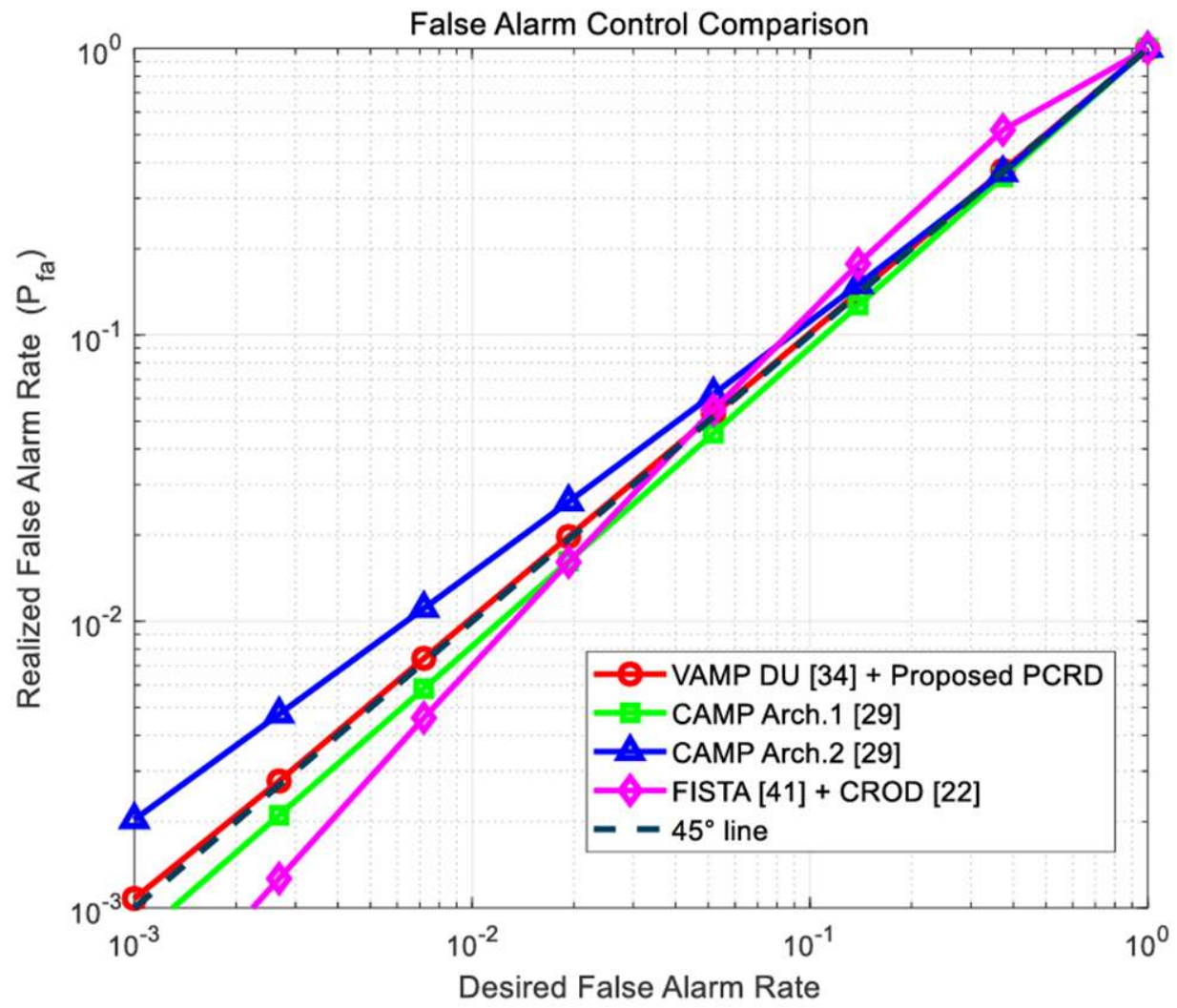

(b)

Fig. 6. Performance comparison of different detection schemes, at SNR = 13 dB: (a) shows that VAMP-DU combined with PCRD achieves better target detection performance; (b) shows that VAMP-DU combined with PCRD exhibits superior false alarm control capability.

PCRD, CROD requires precise knowledge of the AWGN power $\sigma^2$.

From Fig. 5 (b), it can be concluded that the false alarm control performance of VAMP-DU combined with PCRD is superior to that of VAMP-DU combined with CROD. When there is a bias in the estimated power of AWGN, such that $\sigma_{\text{bias}} = 0.9\sigma$ , the accuracy of the false alarm control performance of CROD decreases. The comparison results from Fig. 5 indicate that PCRD is more suitable for VAMP-DU than CROD.

In the following numerical experiments, we set the dimensions of the partial Fourier observation matrix as $M = 600$, $N = 1000$. The parameters for the testing data are set as $\rho_{\min} = \rho_{\max} = 0.03$, $\text{SNR} = 13$ dB, $a_{\min} = a_{\max} = 1.0$. The number of layers for the VAMP-DU is set to $K_{\text{it}} = 7$. The training data parameters for VAMP-DU are set as $\rho_{\min} = 0.01$, $\rho_{\max} = 0.05$, $\text{SNR}_{\min} = 8$ dB, $\text{SNR}_{\max} = 18$ dB, $a_{\min} = 0.7$, $a_{\max} = 1.3$. We conducted 10,000 Monte Carlo experiments to obtain Fig. 6.

Fig. 6 (a) compares the target detection performance achieved by VAMP-DU-based PCRD with the performance of other CS radar detection schemes, including two architectural frameworks of CAMP, FISTA [41] combined with CROD (with precise knowledge of the noise power $\sigma^2$ while other algorithms operate without such pre-acquired information). Fig. 6 (a) illustrates that VAMP-DU-based PCRD achieves the highest target detection rate, superior to that of two architectural frameworks of CAMP and FISTA combined with CROD.

The accuracy of false alarm control is shown in Fig. 6 (b). We observe that VAMP-DU combined with PCRD demonstrates the most accurate false alarm control performance, superior to that of two architectural frameworks of CAMP, and FISTA combined with CROD.

## V. Conclusion

In this study, we prove the Gaussianity of error term in VAMP-DU from the perspective of a more general SE theoretically, thereby allowing the distribution of the test statistic to be determined. On this basis, we propose PCRD based on VAMP-DU to estimate the variance of the error term and implement CS radar detection under a given false alarm rate. We analyze the effectiveness of PCRD to prove the accuracy of PCRD in terms of variance estimation. The VAMP-DU-based PCRD exploits the Gaussian property of the error term in VAMP-DU. This characteristic enables more accurate false alarm control. Furthermore, leveraging the improved recovery accuracy of VAMP-DU, the PCRD further enhances the target detection performance. Numerical simulations validate the Gaussianity of the error term and demonstrate that VAMP-DU-based PCRD can exhibit superior false alarm control capability and achieve enhanced target detection performance.

## Appendix A

We review VAMP-DU in [34], shown as Algorithm 2. VAMP-DU obtains the real-valued sparse vector $\hat{\mathbf{x}}_{\text{R,I}}$ and real-valued pseudo-measurement $\mathbf{r}_{\text{R,I}}$, expressed as:

$$\hat{\mathbf{x}}_{\text{R,I}} = \hat{\mathbf{x}}_{2,K_{\text{it}}}, \mathbf{r}_{\text{R,I}} = \mathbf{r}_{2,K_{\text{it}}}, \tag{68}$$

respectively. Each iteration of VAMP-DU consists of two stages: linear minimum mean squared error (LMMSE) estimation and denoising. The two stages follow similar steps, with the only difference in their respective estimators. The estimator in the LMMSE estimation stage $\mathbf{g}_{1,k}\left(\mathbf{r}_{1k},\gamma_{1k}\right)$ in the $k$-th iteration can be represented as:

$$\mathbf{g}_{1,k}\left(\mathbf{r}_{1k},\gamma_{1k}\right)=\left(\gamma_{wk}\mathbf{A}_{\mathrm{R,I}}^{\top}\mathbf{A}_{\mathrm{R,I}}+\gamma_{1k}\mathbf{I}\right)^{-1}\left(\gamma_{wk}\mathbf{A}_{\mathrm{R,I}}^{\top}\mathbf{y}_{\mathrm{R,I}}+\gamma_{1k}\mathbf{r}_{1k}\right), \tag{69}$$

where $\gamma_{wk}$ represents the variance of AWGN which is modified through backpropagation. The estimator of the denoising stage $\mathbf{g}_{2,k}\left(\mathbf{r}_{2k},\gamma_{2k}\right)$ in the $k$-th iteration represents:

$$\mathbf{g}_{2,k}\left(\mathbf{r}_{2k},\gamma_{2k}\right)=\mathbf{g}_{2}\left(\mathbf{r}_{2k},\lambda_k,\gamma_{2k}\right), \tag{70}$$

where $\lambda_k$ represents the parameter modified through backpropagation and $\mathbf{g}_{2}\left(\mathbf{r}_{2k},\lambda_k,\gamma_{2k}\right)$ represents the element-wise soft-threshold function, which can be represented as [34]:

$$\left[\mathbf{g}_{2}\left(\mathbf{r}_{2k},\lambda_k,\gamma_{2k}\right)\right]_j=\operatorname{sgn}\left(r_{2k,j}\right)\max\left\{\left|r_{2k,j}\right|-\lambda_k/\sqrt{\gamma_{2k}},0\right\}. \tag{71}$$

## APPENDIX B

We propose and prove *Lemma 1.*

*Lemma 1.* Assume that in Recursion (11) and SE equations (17), for all $k$:

(i) The functions $C_i\left(\alpha_i\right)$, $\Gamma_i\left(\gamma_i,\alpha_i\right)$ are continuous at the points $\left(\gamma_i,\alpha_i\right)=\left(\overline{\gamma}_{ik},\overline{\alpha}_{ik}\right)$ for $i=1,2$.

(ii) The function $f_{p,k}\left(p,w^p,\gamma_1\right)$ and its derivative $f'_{p,k}\left(p,w^p,\gamma_1\right)$ are uniformly Lipschitz continuous in $\left(p,w^p\right)$ at $\gamma_1=\overline{\gamma}_{1k}$.

(iii) The function $f_{q,k}\left(q,w^q,\gamma_2\right)$ and its derivative $f'_{q,k}\left(q,w^q,\gamma_2\right)$ are uniformly Lipschitz continuous in $\left(q,w^q\right)$ at $\gamma_2=\overline{\gamma}_{2k}$.

Then,

(a) For any fixed $k$, almost surely the components of $\left(\mathbf{w}^p,\mathbf{p}_0,\ldots,\mathbf{p}_k\right)$ empirically converge as

$$\lim_{N\to\infty}\left\{\left(w_n^p,p_{0n},\ldots,p_{kn}\right)\right\}_{n=1}^{N}\overset{PL(2)}{=}\left(W^p,P_0,\ldots,P_k\right) \tag{72}$$

where $W^p$ is the random variable in the Equation (15) and $\left(P_0,\ldots,P_k\right)$ is a Gaussian random vector of zero mean, independent of $W^p$, with $\mathbb{E}\left[P_k^2\right]=\tau_{1k}$. In addition, we have

$$\lim_{N\to\infty}\left(\alpha_{1k},\gamma_{1k}\right)=\left(\overline{\alpha}_{1k},\overline{\gamma}_{1k}\right) \tag{73}$$

almost surely.

(b) For any fixed $k$, almost surely the components of $\left(\mathbf{w}^q,\mathbf{q}_0,\ldots,\mathbf{q}_k\right)$ empirically converge as

$$\lim_{N\to\infty}\left\{\left(w_n^q,q_{0n},\ldots,q_{kn}\right)\right\}_{n=1}^{N}\overset{PL(2)}{=}\left(W^q,Q_0,\ldots,Q_k\right) \tag{74}$$

where $W^q$ is the random variable in the Equation (15) and $\left(Q_0,\ldots,Q_k\right)$ is a Gaussian random vector of zero mean, independent of $W^q$, with $\mathbb{E}\left[Q_k^2\right]=\tau_{2k}$. In addition, we have

$$\lim_{N\to\infty}\left(\alpha_{2k},\gamma_{2k}\right)=\left(\overline{\alpha}_{2k},\overline{\gamma}_{2k}\right) \tag{75}$$

almost surely.

*Proof.* We use an induction argument. Define the hypothesis $H_{k,\ell}$ as: Part (a) of *Lemma 1* is true up to iteration $k$; and Part (b) of *Lemma 1* is true up to iteration $\ell$. The proof steps are as follows:

1. $H_{0,-1}$ is true;
2. If $H_{k,k-1}$ is true, then so is $H_{k,k}$;
3. If $H_{k,k}$ is true, then so is $H_{k+1,k}$.

The proof of $H_{0,-1}$ is the same as [35, Appendix F]. Next, we show the implication $H_{k,k-1}\Rightarrow H_{k,k}$. The implication

**Algorithm 2** VAMP Deep Unfolding [34]

**Require:** Compressed measurement $\mathbf{y}_{\mathrm{R,I}}\in\mathbb{R}^{2M\times 1}$ and observation matrix $\mathbf{A}_{\mathrm{R,I}}\in\mathbb{R}^{2M\times 2N}$
  Select initial $\mathbf{r}_{10}$ and $\gamma_{10}$
  **for** $k=0,1,\ldots,K_{\mathrm{it}}$ **do**
    // LMMSE estimation
    $\hat{\mathbf{x}}_{1k}=\mathbf{g}_{1,k}\left(\mathbf{r}_{1k},\gamma_{1k}\right)$
    $\alpha_{1k}=\left\langle\mathbf{g}'_{1,k}\left(\mathbf{r}_{1k},\gamma_{1k}\right)\right\rangle$
    $\eta_{1k}=\gamma_{1k}/\alpha_{1k}$
    $\gamma_{2k}=\eta_{1k}-\gamma_{1k}$
    $\mathbf{r}_{2k}=\left(\eta_{1k}\hat{\mathbf{x}}_{1k}-\gamma_{1k}\mathbf{r}_{1k}\right)/\gamma_{2k}$
    // Denoising
    $\hat{\mathbf{x}}_{2k}=\mathbf{g}_{2,k}\left(\mathbf{r}_{2k},\gamma_{2k}\right)$
    $\alpha_{2k}=\left\langle\mathbf{g}'_{2,k}\left(\mathbf{r}_{2k},\gamma_{2k}\right)\right\rangle$
    $\eta_{2k}=\gamma_{2k}/\alpha_{2k}$
    $\gamma_{1,k+1}=\eta_{2k}-\gamma_{2k}$
    $\mathbf{r}_{1,k+1}=\left(\eta_{2k}\hat{\mathbf{x}}_{2k}-\gamma_{2k}\mathbf{r}_{2k}\right)/\gamma_{1,k+1}$
  **end for**
  **Return** $\hat{\mathbf{x}}_{2,K_{\mathrm{it}}},\mathbf{r}_{2,K_{\mathrm{it}}}$

$H_{k,k} \Rightarrow H_{k+1,k}$ can be proven similarly. Let the notation $\mathbf{U}_k = [\mathbf{u}_0 \cdots \mathbf{u}_k] \in \mathbb{R}^{2N\times(k+1)}$ represent the first $k+1$ vectors of $\mathbf{u}_\ell$. And the matrices $\mathbf{V}_{k-1}$, $\mathbf{Q}_{k-1}$ and $\mathbf{P}_k$ are defined similarly. Let $G_k$ be the tuple of the above random matrices,

$$G_k = \{\mathbf{U}_k, \mathbf{P}_k, \mathbf{V}_{k-1}, \mathbf{Q}_{k-1}\}. \tag{76}$$

Then we define the matrices

$$\mathbf{A}_k = [\mathbf{P}_k \quad \mathbf{V}_{k-1}], \quad \mathbf{B}_k = [\mathbf{U}_k \quad \mathbf{Q}_{k-1}]. \tag{77}$$

According to the recursion, the relationship between $\mathbf{A}_k$ and $\mathbf{B}_k$ can be calculated as:

$$\mathbf{A}_k = \mathbf{V}'\mathbf{B}_k. \tag{78}$$

From [35, *Lemma 4*], this conditional distribution of $\mathbf{V}'$ is given by

$$\mathbf{V}'\big|_{G_k} = \mathbf{A}_k\left(\mathbf{A}_k^\top \mathbf{A}_k\right)^{-1}\mathbf{B}_k^\top + \mathbf{U}_{\mathbf{A}_k^\perp}\tilde{\mathbf{V}}\mathbf{U}_{\mathbf{B}_k^\perp}^\top, \tag{79}$$

where $\mathbf{U}_{\mathbf{A}_k^\perp}$ and $\mathbf{U}_{\mathbf{B}_k^\perp}$ are matrices with dimension $2N\times(2N-2k-1)$ whose columns are an orthonormal basis for $\mathrm{Range}(\mathbf{A}_k)^\perp$ and $\mathrm{Range}(\mathbf{B}_k)^\perp$.The matrix $\tilde{\mathbf{V}}$ is Haar distributed and independent of $G_k$. Using Equation (79), $\mathbf{q}_k$ can be calculated as a sum of two terms

$$\mathbf{q}_k = \mathbf{V}'^\top \mathbf{v}_k = \mathbf{q}_k^{\det} + \mathbf{q}_k^{\mathrm{ran}}, \tag{80}$$

where $\mathbf{q}_k^{\det}$ is the deterministic part:

$$\mathbf{q}_k^{\det} = \mathbf{B}_k\left(\mathbf{A}_k^\top \mathbf{A}_k\right)^{-1}\mathbf{A}_k^\top \mathbf{v}_k, \tag{81}$$

and $\mathbf{q}_k^{\mathrm{ran}}$ is the random part:

$$\mathbf{q}_k^{\mathrm{ran}} = \mathbf{U}_{\mathbf{B}_k^\perp}\tilde{\mathbf{V}}^\top \mathbf{U}_{\mathbf{A}_k^\perp}^\top \mathbf{v}_k. \tag{82}$$

According to [35, *Lemma 6*, *Lemma 7*, *Lemma 8*], the components of $\left(\mathbf{w}^q, \mathbf{q}_0, \ldots, \mathbf{q}_k\right)$ almost surely converge empirically as

$$\begin{aligned} &\lim_{N\to\infty}\left\{\left(w_n^q, q_{0n}, \ldots, q_{kn}\right)\right\}_{n=1}^N \\ &\overset{PL(2)}{=} \lim_{N\to\infty}\left\{\left(w_n^q, q_{0n}, \ldots, q_{kn}^{\det} + q_{kn}^{\mathrm{ran}}\right)\right\}_{n=1}^N \\ &\overset{PL(2)}{=} \left(W^q, Q_0, \ldots, Q_k\right), \end{aligned} \tag{83}$$

where $Q_k$ is the random variable, represented as:

$$Q_k = \beta_{k,0}Q_0 + \cdots + \beta_{k,k-1}Q_{k-1} + U_k. \tag{84}$$

Here, $(Q_0, \ldots, Q_{k-1})$ and $U_k$ are Gaussian variables. $(\beta_{k,0}, \ldots, \beta_{k,k-1})$ are coefficients of the linear combination relating $Q_k$ to $(Q_0, \ldots, Q_{k-1})$. Therefore, $(Q_0, \ldots, Q_k)$ follows a Gaussian distribution. This proves Equation (74). Since $\alpha_{1k} \to \overline{\alpha}_{1k}$ and $\gamma_{1k} \to \overline{\gamma}_{1k}$ almost surely, we can derive:

$$\lim_{N\to\infty}\gamma_{2k} = \lim_{N\to\infty}\Gamma_1\left(\gamma_{1k}, \alpha_{1k}\right) = \overline{\gamma}_{2k}. \tag{85}$$

In addition, since $\mathbf{f}'_{q,k}\left(\mathbf{q}, \mathbf{w}^q, \gamma_1\right)$ is Lipschitz continuous in $\left(\mathbf{q}, \mathbf{w}^q\right)$ and continuous in $\gamma_1$, there exists:

$$\begin{aligned} \lim_{N\to\infty}\alpha_{2k} &= \lim_{N\to\infty}\left\langle \mathbf{f}'_{q,k}\left(\mathbf{q}_k, \mathbf{w}^q, \gamma_{1k}\right)\right\rangle \\ &= \mathbb{E}\left[f'_{q,k}\left(Q_k, W^q, \overline{\gamma}_{1k}\right)\right] = \overline{\alpha}_{2k}. \end{aligned} \tag{86}$$

This proves Equation (75). Note that

$$\begin{aligned} \mathbb{E}\left[Q_k^2\right] &= \lim_{N\to\infty}\frac{1}{N}\left\|\mathbf{q}_k\right\|^2 = \lim_{N\to\infty}\frac{1}{N}\left\|\mathbf{v}_k\right\|^2 \\ &= C_1^2\left(\overline{\alpha}_{1k}\right)\left\{\mathbb{E}\left[f_{p,k}^2\left(P_k, W^p, \overline{\gamma}_{1k}\right)\right] - \overline{\alpha}_{1k}^2\tau_{1k}\right\} = \tau_{2k}. \end{aligned} \tag{87}$$

Thus, the implication $H_{k,k-1} \Rightarrow H_{k,k}$ has been proven. The implication $H_{k,k} \Rightarrow H_{k+1,k}$ can be proven similarly. This completes the proof *Lemma 1*.

□